\documentclass[letter,preprint,aps,superscriptaddress,nofootinbib,tightenlines]{revtex4}
\usepackage{amsmath,amssymb}
\usepackage{graphicx}
\usepackage{bm}
\usepackage{comment}

\newcommand{\beq}{\begin{equation}}
\newcommand{\eeq}{\end{equation}}
\newcommand{\bea}{\begin{eqnarray}}
\newcommand{\eea}{\end{eqnarray}}

\def\OMIT#1{{}}

\newcommand{\lsim}{\ \raisebox{-0.7ex}{$\stackrel{\textstyle <}{\sim}$ }}
\newcommand{\gsim}{\ \raisebox{-0.7ex}{$\stackrel{\textstyle >}{\sim}$ }}
\def\pislash{ {\pi\hskip-0.54em /} }

\def\nopi{ {\rm EFT}(\pislash) }
\newcommand{\Choose}[2]{{^{#1}C_{#2}}}

\def\abar{\overline{a}}

\def\sheep{$\overline{\overline{\eta}}_3^L$}
\def\apipi{\overline{a}_{\pi\pi}^{(I=2)}}

\newcount\hour \newcount\hourminute \newcount\minute 
\hour=\time \divide \hour by 60
\hourminute=\hour \multiply \hourminute by 60
\minute=\time \advance \minute by -\hourminute

\begin{document}
\begin{figure}[!t]
\vskip -1.18cm
  \leftline{\includegraphics[width=0.25\textwidth]{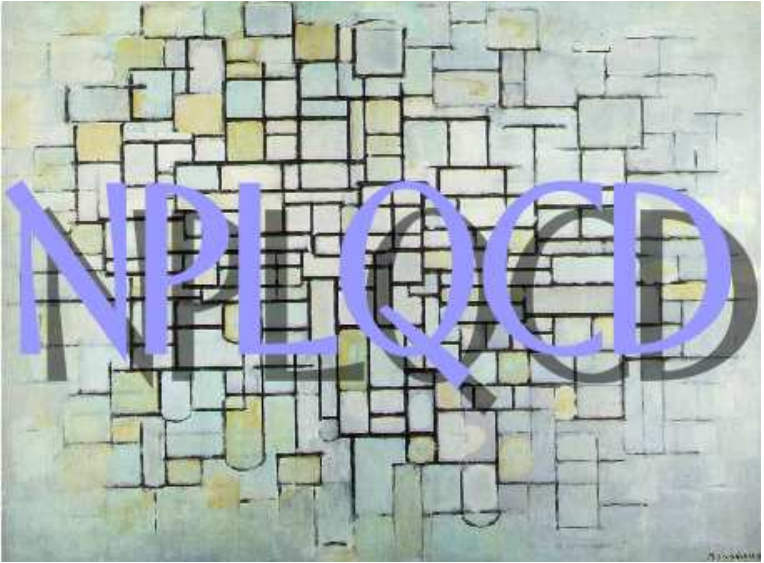}}
\end{figure}

\preprint{
  \vbox{\hbox{} \hbox{UNH-08-01} \hbox{JLAB-THY-08-803}
    \hbox{NT@UW-08-06}
\hbox{LLNL-JRNL-402365}
  }}



\title{Multi-Pion States in Lattice QCD and the Charged-Pion
  Condensate}

\author{William Detmold} \affiliation{Department of
  Physics, University of Washington, Box 351560, Seattle, WA 98195,
  USA.}  
\author{Martin J.~Savage} \affiliation{Department of Physics,
  University of Washington, Box 351560, Seattle, WA 98195, USA.}
\author{Aaron Torok} \affiliation{Department of Physics, University of
  New Hampshire, Durham, NH 03824-3568, USA.}  
\author{Silas R.~Beane}
\affiliation{Department of Physics, University of New Hampshire,
  Durham, NH 03824-3568, USA.}  
\author{Thomas C.~Luu} \affiliation{N
  Division, Lawrence Livermore National Laboratory, Livermore, CA
  94551, USA.}  
\author{Kostas Orginos} \affiliation{Department of
  Physics, College of William and Mary, Williamsburg, VA 23187-8795,
  USA.}  \affiliation{Jefferson Laboratory, 12000 Jefferson Avenue,
  Newport News, VA 23606, USA.}  
\author{Assumpta Parre\~no}
\affiliation{Departament d'Estructura i Constituents de la Mat\`{e}ria
  and Institut de Ci\`encies del Cosmos, Universitat de Barcelona,
  E--08028 Barcelona, Spain.}  \collaboration{ NPLQCD Collaboration }
\noaffiliation \vphantom{}


\begin{abstract}
\noindent 
The ground-state energies of systems containing up to twelve $\pi^+$'s
in a spatial volume $V\sim (2.5~{\rm fm})^3$ are computed in
dynamical, mixed-action lattice QCD at a lattice spacing of
$\sim0.125$~fm for four different values of the light quark masses.
Clean signals are seen for each ground state, allowing for a precise
extraction of both the $\pi^+\pi^+$ scattering length and
$\pi^+\pi^+\pi^+$-interaction from a correlated analysis of systems
containing different numbers of $\pi^+$'s.  This extraction of the
$\pi^+\pi^+$ scattering length is consistent with than that from the
$\pi^+\pi^+$-system alone.  The large number of systems studied here
significantly strengthens the arguments presented in our earlier work
and unambiguously demonstrates the presence of a low energy
$\pi^+\pi^+\pi^+$-interaction.  The equation of state of a $\pi^+$ gas
is investigated using our numerical results and the density dependence
of the isospin chemical potential for these systems agrees well with
the theoretical expectations of leading order chiral perturbation
theory.  The chemical potential is found to receive a substantial
contribution from the $\pi^+\pi^+\pi^+$-interaction at the lighter
pion masses.  An important technical aspect of this work is the
demonstration of the necessity of performing propagator contractions
in greater than double precision to extract the correct results.
\end{abstract}
 \maketitle


%
%
\section{Introduction 
\label{sec:Intro}}

\noindent
Multi-hadron systems, from the deuteron to heavy nuclei to neutron
stars, represent a significant fraction of the universe that we
observe, and for decades the phenomenological study of these systems
defined the field of nuclear physics.  Understanding how nuclei and
nuclear interactions emerge from Quantum Chromodynamics (QCD), the
underlying theory of the strong interaction, is now a central goal of
modern sub-atomic physics.  Since hadrons are bound-states of quarks
and gluons, they do not arise at any finite order in perturbation
theory and a description from QCD has proved elusive. The only known
non-perturbative method that systematically implements QCD from first
principles is its formulation on a discretized space-time, {\it
  lattice QCD}.  While it is still not possible to directly calculate
the properties of even the simplest nucleus from QCD, the tools and
technology are gradually being put in place to make such calculations
possible with lattice QCD in the near future.  The impact of the
successful realization of this goal cannot be overstated. For the
first time, it would allow reliable calculations of
strongly-interacting many-body processes that are not (or are only
poorly) accessible experimentally. Important examples are
hyperon-nucleon interactions that many play significant roles in the
interior of neutron stars. Further, it would enable the exploration of
how the properties of such systems depend upon the fundamental
constants of nature, exposing the (possible) fine-tunings between the
light-quark masses that give rise to the multiple fine-tunings
observed in nuclear physics.

Our current understanding of nuclei requires a small but non-zero
three-nucleon interaction.  In the study of the structure of nuclei,
we now have refined many-body techniques, such as Green's function
Monte-Carlo (GFMC)~\cite{Pieper:2007ax} with which to calculate the
ground states and excited states of light nuclei, with atomic number
$A\lsim 14$.  Using modern nucleon-nucleon potentials that reproduce
all scattering data below inelastic thresholds with $\chi^2/dof\sim
1$, such as ${\rm AV}_{18}$ \cite{Wiringa:1994wb}, one fails, quite
dramatically, to recover the structure of light nuclei.  The inclusion
of three-nucleon interactions greatly improves the predicted structure
of nuclei, but at present, such interactions are difficult to
constrain.  At some point in the future, lattice QCD will be able to
predict the interactions between multiple neutrons (and proton and
mixed proton-neutron systems), bound or unbound in the same way it
will be used to determine the two-body scattering parameters.  A
calculation of the three-neutron interaction, for instance, will be
possible.

As a first step toward the study of nuclei and their interactions
using lattice QCD, in this work we study systems composed of up to
twelve $\pi^+$'s.  These multi-pion systems are conceptually and
computationally the simplest multi-hadron systems that can be
constructed.  In addition to the two-body interactions, there are
expected contributions from multi-body interactions.  Such multi-body
interactions are not forbidden by the symmetries of QCD, and are
expected to be present with a magnitude that can be estimated with
naive dimensional analysis (NDA)~\cite{Manohar:1983md}.  However, they
are qualitatively (and obviously quantitatively) different from
systems involving nucleons.  The lowest-lying continuum state of
multiple pions in a large volume is perturbatively close to each pion
carrying zero-momentum, whereas the wavefunctions of systems of
multiple nucleons are subject to the Pauli exclusion principle.

In a recent letter~\cite{Beane:2007es}, we have reported results of
the first many-pion calculation\footnote{``Many-body'' implies systems
  containing more than two bodies.  When we refer to the pion we will
  mean the $\pi^+$ unless otherwise stated.} in lattice QCD.  We
explored systems containing up to five pions, extracted the
$\pi^+\pi^+$ scattering length, and, for the first time, found
indications of a non-zero renormalization group invariant (RGI)
$\pi^+\pi^+\pi^+$-interaction.  Here we continue our investigations,
presenting more detailed results of the previous studies and extending
our work to systems containing up to twelve charged pions.  We also
use the recently derived expression for the ground-state energy of
$n$-identical bosons in a finite volume at ${\cal O}(L^{-7})$
\cite{Detmold:2008gh} in our analysis, one order beyond that at which
our previous calculations~\cite{Beane:2007es} were analyzed.

Multi-pion systems are of interest in their own right as a strongly
interacting boson gas at finite density and temperature.  Such systems
may be important for the late-time evolution of heavy ion collisions,
such as those at RHIC and also in the interior of neutron
stars~\cite{Sawyer:1972cq,Scalapino:1972fu}.  During the last several
years there have been a number of theoretical explorations of pionic
systems at finite isospin chemical potential, with or without a finite
baryon number chemical potential.  Leading order (LO) chiral
perturbation theory ($\chi$PT) has been used to study the vacuum
realignment that takes place in the presence of an isospin chemical
potential that exceeds the mass of the pion, leading to a charged pion
condensate, and excitations about this realigned vacuum.  One of the
important results of this present work is the calculation of the
isospin chemical potential as a function of the isospin density.  Our
results are in good agreement with the LO $\chi$PT result, and
further, demonstrate the sizable contribution from multi-pion
interactions even at moderate densities.

The structure of this paper is as follows. In Section
\ref{sec:multi-meson-energies} we review the theoretical expectations
for the ground state energy of multi-pion systems at finite volume and
discuss methods for extracting their interactions.  In
Section~\ref{sec:lattice-techniques} we provide details of our lattice
QCD measurements and analysis and in Sections
\ref{sec:correlation-energies} and \ref{sec:threebody} we present the
main results of our calculations.  In Section \ref{sec:chemical} we
discuss the implications of our results for the equation of state of
the pionic gas, its isospin chemical-potential and its pressure.
Finally in Section \ref{sec:disc} we discuss the results in a global
context and conclude. Certain technical details of contractions and
numerical implementation are relegated to the Appendices.

\section{Multi-Meson Energies : Isolating the Two- and Three-Body Interactions}
\label{sec:multi-meson-energies}

\noindent
In recent works \cite{Beane:2007qr,Detmold:2008gh,Tan:2007bg}, the
analytic volume dependence of the energy of $n$ identical bosons in a
periodic volume has been computed to ${\cal O}(L^{-7})$, extending the
classic results of Bogoliubov \cite{Bog} and Lee, Huang and Yang
\cite{Lee:1957}.  The resulting shift in energy of $n$ particles of
mass $M$ due to their interactions is \cite{Detmold:2008gh}
\begin{eqnarray}
 \Delta E_n &=&
  \frac{4\pi\, \abar}{M\,L^3}\Choose{n}{2}\Bigg\{1
-\left(\frac{\abar}{\pi\,L}\right){\cal I}
+\left(\frac{\abar}{\pi\,L}\right)^2\left[{\cal I}^2+(2n-5){\cal J}\right]
\nonumber 
\\&&\hspace*{2cm}
-
\left(\frac{\abar}{\pi\,L}\right)^3\Big[{\cal I}^3 + (2 n-7)
  {\cal I}{\cal J} + \left(5 n^2-41 n+63\right){\cal K}\Big]
\nonumber
\\&&\hspace*{2cm}
+
\left(\frac{\abar}{\pi\,L}\right)^4\Big[
{\cal I}^4 - 6 {\cal I}^2 {\cal J} + (4 + n - n^2){\cal J}^2 
+ 4 (27-15 n + n^2) {\cal I} \ {\cal K}
\nonumber\\
&&\hspace*{4cm}
+(14 n^3-227 n^2+919 n-1043) {\cal L}\ 
\Big]
\Bigg\}
\nonumber\\
&&
+\ \Choose{n}{3}\left[\ 
{192 \ \abar^5\over M\pi^3 L^7} \left( {\cal T}_0\ +\ {\cal T}_1\ n \right)
\ +\ 
{6\pi \abar^3\over M^3 L^7}\ (n+3)\ {\cal I}\ 
\right]
\nonumber\\
&&
+\ \Choose{n}{3} \ {1\over L^6}\ \overline{\overline{\eta}}_3^L\ 
\ \ + \ {\cal O}\left(L^{-8}\right)
\ \ \ \ ,
\label{eq:energyshift}
\end{eqnarray}
where the parameter $\abar$ is related to the scattering
length\footnote{In this work we use the Nuclear Physics sign
  convention for the scattering length, which is opposite to that of
  the Particle Physics sign convention.  In this convention, the
  $\pi^+\pi^+$ scattering length is positive.}, $a$, and the effective
range, $r$, by
\begin{eqnarray}
a
& = & 
\overline{a}\ -\ {2\pi\over L^3} \overline{a}^3 r \left(\ 1 \ -\
  \left( {\overline{a}\over\pi L}\right)\ {\cal I} \right)\ \ .
\label{eq:aabar}
\end{eqnarray}
The geometric constants that enter into eq.~(\ref{eq:energyshift}) are
\begin{align} 
  &{\cal I}\ =\ -8.9136329\,, &{\cal J}\ =\ 16.532316\,, \qquad\qquad
  {\cal K}\ = \ 8.4019240\,,
  \nonumber\\
  &{\cal L}\ = \ 6.9458079\,, &{\cal T}_0\ = -4116.2338\,,
  \qquad\qquad {\cal T}_1\ = \ 450.6392\,, &
\label{eq:sums}
\end{align}
and $^nC_m=n!/m!/(n-m)!$.  The three-body contribution to the
energy-shift given in eq.~(\ref{eq:energyshift}) is represented by the
parameter $\overline{\overline{\eta}}_3^L$, which is a combination of
the volume-dependent, renormalization group invariant quantity,
$\overline{\eta}_3^L$, and contributions from the two-body scattering
length and effective range,
\begin{eqnarray}
\overline{\overline{\eta}}_3^L & = & \overline{\eta}_3^L  \left(\ 1 \ -\
  6  \left({\overline{a}\over\pi L}\right)\ {\cal I} \right)
\ +\ {72\pi \overline{a}^4 r\over M L} \ {\cal I}
\ \ \ ,
\label{eq:eta3barbar}
\end{eqnarray}
where
\begin{align} 
  \overline{\eta}_3^L & = \eta_3(\mu)\ +\ {64\pi a^4\over
    M}\left(3\sqrt{3}-4\pi\right)\ \log\left(\mu L\right)\ -\ {96
    a^4\over\pi^2 M} {\cal S}_{\rm MS} \ \ \ .
\label{eq:etathreebar}
\end{align}
The quantity $\eta_3(\mu)$ is the coefficient of the three-$\pi^+$
interaction that appears in the effective Hamiltonian density
describing the system \cite{Detmold:2008gh}. It is renormalization
scale, $\mu$, dependent.  The quantity ${\cal S}$ is renormalization
scheme dependent and we give its value in the minimal subtraction (MS)
scheme, ${\cal S}_{\rm MS}\ = \ -185.12506$.

For $n=2$, the last two terms in eq.~(\ref{eq:energyshift}) vanish and
the remaining terms constitute the small $\abar/L$ expansion of the
exact eigenvalue equation derived by L\"usher
\cite{Luscher:1986pf,Luscher:1990ux}:
\begin{eqnarray}
p\cot\delta(p) \ =\ {1\over \pi L}\ {\bf
  S}\left(\,\left(\frac{p L}{2\pi}\right)^2\,\right)
\ \ ,
\label{eq:energies}
\end{eqnarray}
which is valid below the inelastic threshold.  Below this threshold
$p\cot\delta(p)$ is the real part of the inverse scattering amplitude.
The regulated three-dimensional sum is~\cite{Beane:2003da}
\begin{eqnarray}
{\bf S}\left(\, x \, \right)\ \equiv \ \sum_{\bf j}^{ |{\bf j}|<\Lambda}
{1\over |{\bf j}|^2-x}\ -\  {4 \pi \Lambda}
\ \ \  ,
\label{eq:Sdefined}
\end{eqnarray}
where the summation is over all triplets of integers ${\bf j}$ such
that $|{\bf j}| < \Lambda$ and the limit $\Lambda\rightarrow\infty$ is
implicit.  For $n=3$, eq.~(\ref{eq:energyshift}) reproduces the shift
of the ground state energy of three identical bosons that was recently
calculated by Tan \cite{Tan:2007bg}.

Naively, one might have expected to be able to determine the two-body
effective range parameter, $r$, by calculating the energies of systems
with different numbers of pions.  However, given the scattering
parameter redefinitions of eq.~(\ref{eq:aabar}), this is clearly not
possible.\footnote{Writing $-1/p\cot\delta = \abar = a + a^2 r p^2/2 +
  ...$, and evaluating it at the shifted energy of two particles in
  the volume at LO in the volume expansion, gives $\abar = a + 2\pi
  a^3 r/L^3 + ...$.  }  Instead, the effective range will be extracted
by calculating the energies of the excited states of, for example, the
$n=2$ pion system at finite volume, the lowest lying of which is
perturbatively close to the state in which the pions both carry
(back-to-back) one unit of lattice momentum, $2\pi/L$.

It is useful to form various combinations of the many-body energies in
order to isolate or eliminate the various contributions from two-body
or three-body interactions.  Further, important checks can be made
regarding the convergence of the large volume expansion, as more
particles are added into the volume.  In particular, the combinations
involving systems with $n$, $m$ and two bodies
\begin{eqnarray}
  \label{eq:vanishesL6}
\zeta_{n,m}^{(6)} & = &   1-\frac{(m-2)}{(m-n)\Choose{n}{2}}\left(\frac{\Delta E_n}{\Delta E_2}
-\frac{\Choose{n}{3}}{\Choose{m}{3}}\frac{\Delta E_m}{\Delta E_2}\right)
+  5 (4 - 2 m - 2 n + m n) {\cal K}\left(\frac{L^2 M\Delta E_2}{4
    \pi^2}\right)^3
\nonumber \\
\end{eqnarray}
vanish at order ${\cal O}(L^{-6})$ while
\begin{eqnarray}
  \label{eq:vanishesL7}
\zeta_{n,m}^{(7)} & = &  1-\frac{(m-2)}{(m-n)\Choose{n}{2}}\left(\frac{\Delta E_n}{\Delta E_2}
-\frac{\Choose{n}{3}}{\Choose{m}{3}}\frac{\Delta E_m}{\Delta E_2}\right)
+  5 (4 - 2 m - 2 n + m n) {\cal K}\left(\frac{L^2 M\Delta E_2}{4 \pi^2}\right)^3
\nonumber\\
&&-{(m-2)(n-2)\over 32\pi^2} {\cal I} \left(L \ \Delta E_2\right)^2
\nonumber\\
&&\ +\ 
{(m-2)(n-2)\over 256\pi^8} \left(M L ^2  \Delta E_2\right)^4 
\left( {\cal J}^2 + 16 {\cal I} {\cal K} + (199 -14 m -14 n) {\cal L} - 16
  {\cal T}_1 \right)
\ ,
\end{eqnarray}
vanishes at order ${\cal O}(L^{-7})$ for any $n$, $m$.

The three-body interaction, \sheep\ , can be eliminated by forming
combinations of the many-body energies, allowing for various
determinations of $\abar$.  One such combination is
\begin{eqnarray}
  \label{eq:abarisolation}
&& {3 L^3 M \over\pi}
  \frac{(\ \Choose{m}{3}{\Delta E}_n  - \Choose{n}{3}{\Delta E}_m\ )
}{ n\  m\  (m-1) (n-1) (m - n)} 
\ =\ 
\overline{a} \Bigg\{
1\ - \ 
\frac{\overline{a}}{\pi \,L}{\cal I}
\nonumber
 \\
&&
\qquad\qquad
+\ 
\left(\frac{\overline{a}}{\pi \,L}\right)^2\left[{\cal I}^2-{\cal J}
\ +{ {\cal I}\over M^2 L^2} ( m+n-2-m n)
\right]
\nonumber
 \\
&&
\qquad\qquad
+\left(\frac{\overline{a}}{\pi \,L}\right)^3
\left[-{\cal I}^3 + 3
   {\cal IJ} +  (19 + 5 m \,n - 10 (n+m)){\cal K}\right]
\nonumber
 \\
&&
\qquad\qquad
+\left(\frac{\overline{a}}{\pi \,L}\right)^4 \Big[
{\cal I}^4-6 {\cal  J} {\cal I}^2 - 4  {\cal I} {\cal K} (m (n-2)-2 n+3)
\nonumber
 \\
&&
\qquad\qquad\qquad
+{\cal J}^2 (m
(n-2)-2 n+6)-16 (m-2) (n-2) {\cal T}_{1}
 \\
&&
\qquad\qquad\qquad
+{\cal L} \left(-14 (n-2)
   m^2-(n-2) (14 n-227) m+28 n^2-454 n+795\right)\Big]
\Bigg\}
\ \ ,
\nonumber
\end{eqnarray}
which allows for the scattering length to be extracted at N$^3$LO
(omitting the last set of square brackets) and N$^4$LO in the large
volume expansion.

Similarly, the three-body interaction can be isolated from
combinations of the many-body energies.  One such combination formed
from the $n$-body and the two-body energies is
\begin{eqnarray}
  \label{eq:eta3barbarisolation}
 \overline{  \overline{\eta}}_3^{L} 
&=&
L^6 {1\over\ \Choose{n}{3} }\ 
\left[\ \Delta E_n - \Choose{n}{2}  \ \Delta E_2 \right]
\ -\ 
L\  {3\ M^2\  {\cal J} \over 8\ \pi^4\ } \left( \Delta E_2\ L^3\right)^3
\nonumber\\
&& -\ 
\ {3 \ M^3 \ ( 4 \ {\cal I} \ {\cal J} \ +\  {\cal K}\ (31 - 5 n ))\over
  64\ \pi^6\ }\left( \Delta E_2\ L^3\right)^4
\nonumber\\
&& -\ 
{1\over L}\  { 3\ M^4\over 256\ \pi^8\ }
\left( 4\  {\cal I}^2\  {\cal J} + {\cal L} \ (521 - 199 n + 14 n^2) + 16 {\cal T}_0 + 16 n {\cal T}_1  
\right.\nonumber\\
&&\left. \qquad\qquad \qquad\qquad
- 8\ {\cal I} \ {\cal K}\  (2 n - 9) - {\cal J}^2 \ ( n - 5)
\right)\left(\ \Delta E_2\ L^3\right)^5
\nonumber\\
&& -\ 
{1\over L}\ 
{3 \ {\cal I}\  (n+3)\over 32 \pi^2 }  \left(\ \Delta E_2 L^3\right)^3
\ \ \ ,
\end{eqnarray}
which allows for $ \overline{ \overline{\eta}}_3^{L}$ to extracted at
two orders in the expansion.  This, of course, can be
straightforwardly generalized to other combinations of energies that
may or may not include $\Delta E_2$.

The ground-state energies at finite volume, given in
eq.~(\ref{eq:energyshift}), have been computed in non-relativistic
quantum mechanics, with relativistic effects added
perturbatively~\cite{Detmold:2008gh}.  They are given in terms of the
scattering parameters, and the three-body interaction.  For pionic
systems in particular, it is natural to ask about the role of chiral
perturbation theory in such a calculation.  In $\chi$PT, the expansion
parameters for this system (in the $p$-regime) are $p/\Lambda_\chi$
and $m_\pi/\Lambda_\chi$, where $\Lambda_\chi\sim 4\pi f_\pi$ is the
chiral symmetry breaking scale.  The $\chi$PT expression for the
ground state energy of $n$-$\pi$'s in a finite volume (which remain to
be determined) will be a dual expansion in $2\pi/(\Lambda_\chi L)$ and
$m_\pi/\Lambda_\chi$.  As such, in order to extract the three-body
interaction that first enters at ${\cal O}(L^{-6})$, the calculation
in $\chi$PT would need to be performed at N$^3$LO.  While providing
the complete quark-mass dependence at this order, such a calculation
would involve the evaluation of three-loop diagrams, and contributions
from the relevant N$^3$LO counterterms.  Organizing the perturbative
expansion using non-relativistic effective field theory, $\nopi$ (see
e.g. Refs.~\cite{Kaplan:1998we,Chen:1999tn}), greatly simplifies the
result, but does so at the expense of ignorance of the quark
mass-dependence of the $\nopi$ parameters without further calculation
(such as the quark mass dependence of the scattering length).

It is important to notice that for the systems containing a large
enough number of pions, the energy-shift of the ground state exceeds
that required to pair-produce pions.  The leading relativistic effects
that are included in the analytic expression for the energy-shift
include only the single particle relativistic kinematics and
corrections to the two-pion scattering amplitude, they do not include
contributions due to inelasticities, including pair-production.
However, we conclude that such effects make a small contribution to
the ground-state energy as we find no evidence for deviations from
eq.~(\ref{eq:energyshift}). Nonetheless, this aspect of these systems
must be explored further.

%
%
\section{Methodology and Details of the Lattice Calculation}
\label{sec:lattice-techniques}

\subsection{Lattice configurations and quark propagators}
\label{sec:lattice-details}

\noindent
The results of the numerical computations presented in this paper were
obtained using the mixed-action lattice QCD scheme developed by
LHPC~\cite{Renner:2004ck,Edwards:2005kw} and are based on the coarse
MILC lattice configurations~\cite{Bernard:2001av}. These lattices have
a lattice spacing of $b\sim 0.125$~fm, and a spatial extent of $L\sim
2.5$~fm. They were generated using the
asqtad-improved~\cite{Orginos:1999cr,Orginos:1998ue} staggered
formulation of lattice fermions, taking the fourth root of the fermion
determinant, and the one-loop, tadpole-improved Symanzik gauge
action~\cite{Alford:1995hw}.  Herein, we assume that the ``fourth-root
trick''\footnote{For an introduction to staggered fermions and the
  fourth-root trick, see Ref.~\cite{degrandANDdetar}. For the most
  recent discussions of the topic, see
  Ref.~\cite{Durr:2004as,Durr:2004ta,Creutz:2006ys,Bernard:2006zw,
    Bernard:2006vv,Creutz:2007nv,Bernard:2006ee,Bernard:2006qt,
    Creutz:2007yg,Creutz:2007pr,Durr:2006ze,Hasenfratz:2006nw,
    Shamir:2006nj,Sharpe:2006re}} recovers the correct continuum limit
of QCD. These ensembles of configurations have a fixed (almost
physical) strange quark mass while the degenerate light quarks were
varied over a range of masses; see Table~\ref{tab:MILCcnfs} and
Refs.~\cite{Beane:2006mx,Beane:2006pt,Beane:2006fk,Beane:2006kx,Beane:2006gf}
for details.

Based on these configurations, valence quark propagators using the
domain-wall (DW) formulation of the lattice fermion
action~\cite{Kaplan:1992bt,Shamir:1992im,Shamir:1993zy,Shamir:1998ww,
  Furman:1994ky} were computed from smeared sources on each
gauge-field configuration.  Hyper-cubic (HYP)
smearing~\cite{Hasenfratz:2001hp,DeGrand:2002vu,DeGrand:2003in,Durr:2004as}
was applied to the gauge links used in the domain-wall fermion action
to improve chiral symmetry, and in calculating the quark propagators,
Dirichlet boundary conditions were imposed to reduce the original
temporal extent of 64 down to 32.  This procedure is optimized for
nucleon physics and indeed leads to minimal degradation of a nucleon
signal, however it does limit the number of time slices available for
fitting meson properties in which the ratio of signal to noise remains
constant in time. Further details about the mixed-action scheme can be
found in Refs.~\cite{Beane:2006gf,Beane:2006gj}.  A summary of the
lattice parameters and resources used in this work is given in
Table~\ref{tab:MILCcnfs}.  In order to generate large statistics on
the existing MILC configurations, multiple propagators from sources
displaced both temporally and spatially on the lattice were computed.
%
%
\begin{table}[t]
 \caption{The parameters of the MILC gauge configurations and
   domain-wall propagators used in this work. The subscript $l$
   denotes light quark (up and down), and  $s$ denotes the strange
   quark. The superscript $dwf$ denotes the bare-quark mass for the
   domain-wall fermion propagator calculation. The last column is the 
   number of configurations times the number of sources per
   configuration. Throughout the paper we will use the pion mass to
   refer to the ensembles. 
}
\label{tab:MILCcnfs}
\begin{ruledtabular}
\begin{tabular}{ccccccc}
 Ensemble        
&  $b m_l$ &  $b m_s$ & $b m^{dwf}_l$ & $ b m^{dwf}_s $ & $m_\pi$ [ MeV ] & \# of propagators   \\
\hline 
2064f21b676m007m050 &  0.007 & 0.050 & 0.0081 & 0.081  & $\sim 291$ & 1039\ $\times$\ 24 \\
2064f21b676m010m050 &  0.010 & 0.050 & 0.0138 & 0.081  & $\sim 352$ & 769\ $\times$\ 24 \\
2064f21b679m020m050 &  0.020 & 0.050 & 0.0313 & 0.081  & $\sim 491$ & 486\ $\times$\ 24 \\
2064f21b681m030m050 &  0.030 & 0.050 & 0.0478 & 0.081  & $\sim 591$ & 564\ $\times$\ 23 \\
\end{tabular}
\end{ruledtabular}
\end{table}

In the continuum chiral limit the $n_f=2$ staggered action has an
$SU(8)_L\otimes SU(8)_R\otimes U(1)_V$ chiral symmetry due to the
four-fold taste degeneracy of each flavor, and each pion has 15
degenerate partners.  At finite lattice spacing, this symmetry is
broken and the taste multiplets are no longer degenerate, but have
splittings that are ${\cal O}(\alpha^2 b^2)$ for the asqtad staggered
action.  When determining the mass of the DW valence quarks there is
an ambiguity due to the non-degeneracy of the 16 staggered bosons
associated with each pion.  One could choose to match to the
taste-singlet meson or to any of the mesons that become degenerate in
the continuum limit.  The choice of tuning to the lightest taste of
staggered meson mass, as opposed to one of the other tastes, provides
for the ``most chiral'' domain-wall mesons and therefore reduces the
uncertainty in extrapolating to the physical point.  The mass
splitting between the domain-wall mesons and the staggered
taste-identity mesons, which characterizes the unitarity violations
present in the calculation, is then given
by~\cite{Aubin:2004fs,Bernard:2006wx}
\begin{align}
  b^2 m_{\pi_I}^2 - b^2 m_{\pi_{dwf}}^2 &= 0.0769(22) \ \ \ .
\end{align}

\begin{table}[!t]
\caption{The mass and decay constant of the $\pi^+$, and the energy-shift and
  scattering parameters in the  $\pi^+\pi^+$ system calculated
  previously~\cite{Beane:2007xs}. The first uncertainties are statistical,
  the second uncertainties are systematic uncertainties due to fitting
  and the third 
  uncertainty, when present, is a comprehensive systematic
  uncertainty~\cite{Beane:2007xs}. $l^{(I=2)}_{\pi\pi}$ is the
  one-loop counterterm in 
  $\chi$PT contributing to $\pi^+\pi^+$ scattering, and $\delta$ is the
  phase-shift. 
Recall that we are using the nuclear physics convention for the sign of the
scattering length.
} 
\label{table:su2fits}
\resizebox{!}{2.28cm}{
\begin{tabular}{@{}|c | c | c | c | c |}
\hline
\  Quantity \ & 
\ \ \ $\qquad m_\pi\sim 291\  {\rm MeV}\qquad$\ \ \ & 
\ \ \ $\qquad m_\pi\sim 352\ {\rm MeV}\qquad$\ \ \ & 
\ \ \ $\qquad m_\pi\sim 491\ {\rm MeV}\qquad$\ \ \ & 
\ \ \ $\qquad m_\pi\sim 591\  {\rm MeV}\qquad$\ \  \  \\
\hline
\ Fit Range  \ & 
$\qquad 8-12\qquad$ & 
$\qquad 8-13 \qquad$ & 
$\qquad 7-13\qquad$  & 
$\qquad 9-12\qquad$ \\
\hline
$m_\pi$ (l.u.) & $0.18454(58)(51)$ & 
        $0.22294(31)(09)$ & 
        $0.31132(28)(21)$ & 
        $0.37407(49)(12)$\\
$f_\pi$ (l.u.) & $0.09273(29)(42)$ & 
        $0.09597(16)(10)$ & 
        $0.10179(12)(28)$ & 
        $0.10759(28)(17)$\\
$m_\pi/f_\pi$ & $1.990(11)(14)$ & 
        $2.3230(57)(30)$ & 
        $3.0585(49)(95)$ & 
        $3.4758(98)(60)$\\
\hline
\ Fit Range  \ & 
$\qquad 11-15\qquad$ & 
$\qquad 9-15 \qquad$& 
$\qquad 10-15\qquad$  & 
$\qquad 12-17\qquad$  \\
\hline
$\Delta E_2$ (l.u.) & $0.00779(47)(14)$  & 
        $0.00745(20)(07)$ & 
        $0.00678(18)(20)$ & 
        $0.00627(23)(10)$\\
$m_\pi \overline{a}_{\pi\pi}^{(I=2)}$ ($b\ne 0$) & 
        $0.1458(78)(25)(14)$ & 
        $0.2061(49)(17)(20)$ & 
        $0.3540(68)(89)(35)$  & 
        $0.465(14)(06)(05)$\\
$l^{(I=2)}_{\pi\pi}$\ ($b\ne 0$) & 
        $6.1(1.9)(0.7)(0.4)$ & 
        $5.23(68)(24)(28)$ & 
        $6.53(32)(42)(16)$ &
        $6.90(40)(18)(13)$\\
$\delta\ (b\ne 0) ({\rm degrees}) $ 
& $-1.71(14)(04)$
& $-2.181(81)(28)$
& $-3.01(09)(12)$
& $-3.46(17)(07)$\\
$|{\bf p}|/m_\pi$ 
& $0.2032(60)(18)$
& $0.1836(25)(09)$
& $0.1480(17)(23)$
& $0.1298(24)(10)$\\
\hline
\end{tabular}}
\end{table}
Simple properties of the $\pi^+$ have been computed to high precision
on the ensembles of coarse MILC lattices that are used in this work,
both with staggered valence quarks and domain-wall valence quarks.
Further, the energies of the lowest-lying $2$-$\pi^+$ states, which
lead directly to the scattering lengths using L\"uscher's method, have
been computed relatively precisely~\cite{Beane:2007xs}.  The results
of the previous mixed-action calculations on these lattice
ensembles~\cite{Beane:2007xs} are shown in
Table~\ref{table:su2fits}\footnote{Until this point the two-body
  scattering length for a generic system has been denoted by $\abar$.
  For the $\pi^+\pi^+$ system, we denote the scattering length by
  $\apipi$.  }.

The results of the present calculation are presented in lattice units
(l.u.), or in terms of dimensionless quantities such as $m_\pi/f_\pi$
which eliminates the requirement of scale setting.  They are performed
only at one lattice spacing, due to limited computer time, and as a
result the continuum limit cannot be determined.  Unlike the two meson
system, for which mixed-action chiral perturbation theory
(MA$\chi$PT)~\cite{Bar:2005tu,Chen:2006wf,Chen:2007ug} has been used
to include the leading order effects of the finite lattice spacing,
MA$\chi$PT calculations have not yet been performed for the
multi-$\pi^+$ systems, and therefore the leading lattice spacing
artifacts in these calculations cannot be removed at present.  The
lattice spacing artifacts are assumed to be small, occurring at ${\cal
  O}(b^2)$, but a systematic study must be performed in the future.

\subsection{Correlation functions}
\label{sec:corr-funct}

\noindent
In this work we determine the $\pi^+\pi^+$ and $\pi^+\pi^+\pi^+$
interactions from the ground-state energy of $n<13$ $\pi^+$'s (isospin
stretched states).  By working in the $m_u=m_d$ limit and restricting
the calculation to states of maximal isospin, only the simplest sets
of propagator contractions are required to be performed (i.e. no
disconnected diagrams) in order to form the correlation functions from
which the ground-state energies are extracted.

\begin{figure}[!t]
  \centering
  \includegraphics[width=0.2\columnwidth]{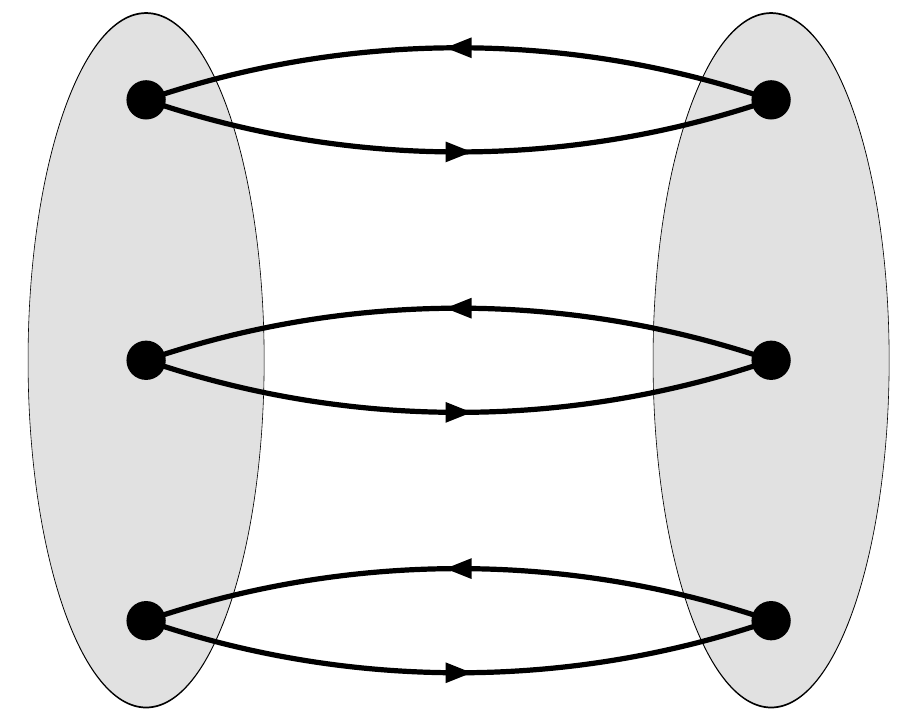}
  \qquad
  \includegraphics[width=0.2\columnwidth]{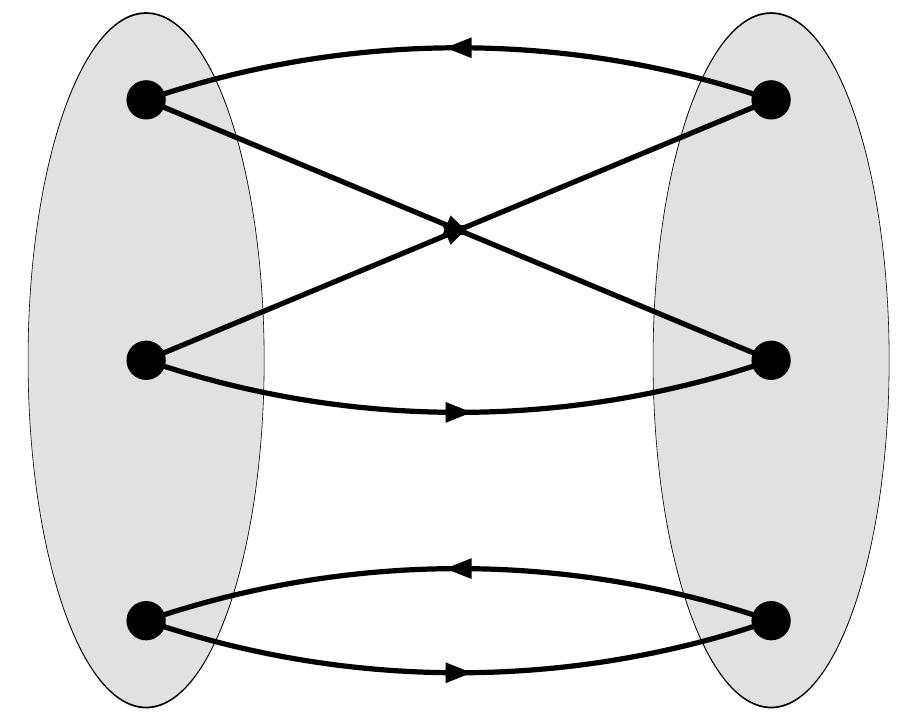}
  \qquad
  \includegraphics[width=0.2\columnwidth]{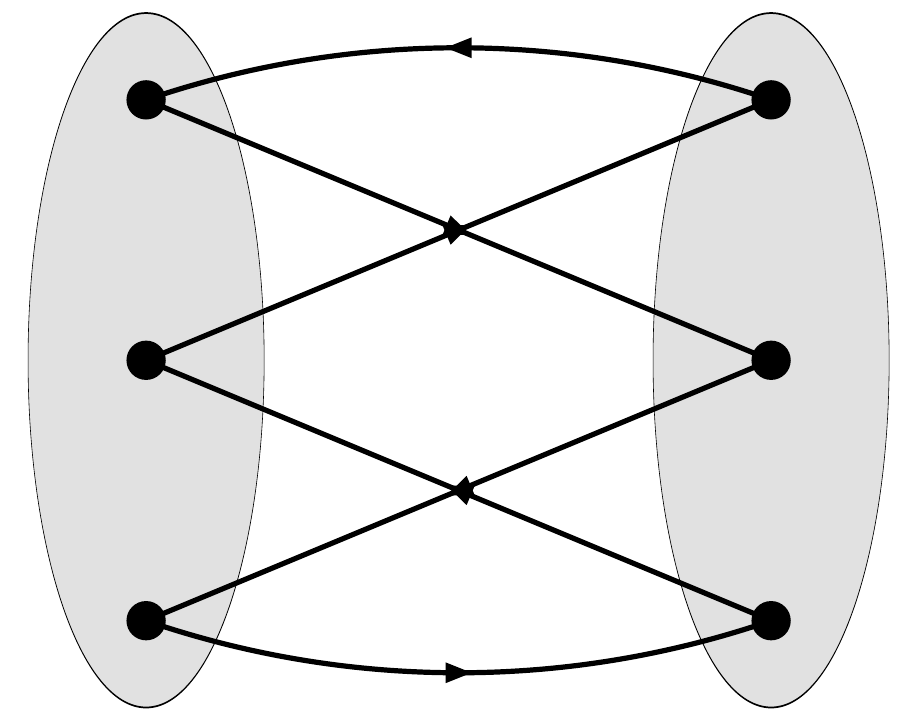}
  \qquad
  \includegraphics[width=0.2\columnwidth]{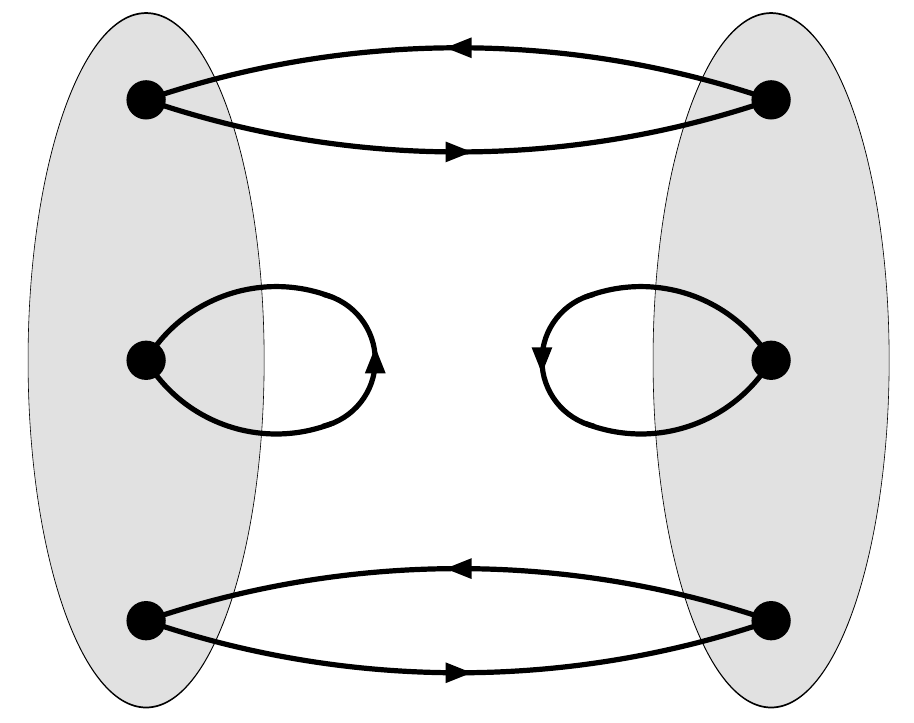}
  \\
  (a)\hspace*{3.55cm} (b)\hspace*{3.55cm} (c)\hspace*{3.55cm} (d)
  \caption{Graphical representation of the contractions for three
    pions with $I_z=3$, (a,b,c). By restricting to the maximal
    isospin, computationally demanding contractions such as the type
    shown in (d) are eliminated.}
  \label{fig:3+5pi_contractions}
\end{figure}

Naively, there are $(n!)^2$ contractions (for large $n$ this behaves
as $\sim (2 n+{1\over 3}) \pi e^{2 n (\log n -1)}$) contributing to
the correlation function of $n$-$\pi^+$'s,
\begin{eqnarray}
C_n(t) 
 & \propto & \langle 
\left(\sum_{\bf x} \pi^-({\bf x},t)
\right)^n
\left( 
\phantom{\sum_x\hskip -0.2in}
\pi^+({\bf 0},0)
\right)^n
\rangle\
 \ \ ,
\label{eq:Cnfun}
\end{eqnarray}
where $\pi^+({\bf x},t)=\overline{u}({\bf x},t)\gamma_5 d({\bf x},t)$.
However, this correlation function can be written as~\footnote{ We
  thank David Kaplan and Michael Endres for discussions on this topic.
  For a general approach to evaluating contractions involving a large
  number of fermions, see Ref.~\cite{OFT}.  }
\begin{eqnarray}
C_n(t) 
 & \propto & 
\langle \ \left(\ \overline{\eta} \Pi \eta\ \right)^n \ \rangle
 \ \ ,
\label{eq:Cnfungrassman}
\end{eqnarray}
where
\begin{eqnarray}
  \Pi
&=& \sum_{\bf x} \ S({\bf x},t;0,0) \   S^\dagger({\bf x},t;0,0)
\ \ \  ,
\label{eq:PiDef}
\end{eqnarray}
and $S({\bf x},t;0,0)$ is a light-quark propagator.  The object
(block) $\Pi$ is a $12\times 12$ (4-spin and 3 color) bosonic
time-dependent matrix, and $\eta_\alpha$ is a twelve component
Grassmann variable.  Using
\begin{eqnarray}
\langle \overline{\eta}^{\alpha_1} \overline{\eta}^{\alpha_2}... 
\overline{\eta}^{\alpha_n} 
\eta_{\beta_1} \eta_{\beta_2} ... \eta_{\beta_n} \rangle
& \propto & 
\varepsilon^{\alpha_1\alpha_2..\alpha_n\xi_1..\xi_{12-n}}\ 
\varepsilon_{\beta_1\beta_2..\beta _n\xi_1..\xi_{12-n}}\ 
\ \ \  ,
\label{eq:GrassCon}
\end{eqnarray}
leads to correlation functions
\begin{eqnarray}
C_n(t) 
 & = & 
\varepsilon^{\alpha_1\alpha_2..\alpha_n\xi_1..\xi_{12-n}}\ 
\varepsilon_{\beta_1\beta_2..\beta _n\xi_1..\xi_{12-n}}\ 
\left(\Pi\right)_{\alpha_1}^{\beta_1} \left(\Pi\right)_{\alpha_2}^{\beta_2} 
.. \left(\Pi\right)_{\alpha_n}^{\beta_n} 
\ \ \  .
\label{eq:Cnepep}
\end{eqnarray}
While correct, further simplifications are possible.  Let us recall
that for an arbitrary $12\times 12$ matrix, $A$,
\begin{eqnarray}
\det\left(1+\lambda A\right) & = & 
{1\over 12!}\ 
\varepsilon^{\alpha_1\alpha_2..\alpha_{12}}\ 
\varepsilon_{\beta_1\beta_2..\beta_{12}}\ 
\left(1+\lambda A\right)_{\alpha_1}^{\beta_1} 
\left(1+\lambda A\right)_{\alpha_2}^{\beta_2} 
\ldots\left(1+\lambda A\right)_{\alpha_{12}}^{\beta_{12}} 
\nonumber\\
& = & 
{1\over 12! }\ \left[\ 
\varepsilon^{\alpha_1\alpha_2..\alpha_{12}}\ 
\varepsilon_{\alpha_1\alpha_2..\alpha_{12}}\ 
\ +\ 
\lambda \ \Choose{12}{1}\ 
\varepsilon^{\alpha_1\alpha_2..\alpha_{12}}\ 
\varepsilon_{\beta_1\alpha_2..\alpha_{12}}\ 
\left(\ A\ \right)_{\alpha_1}^{\beta_1}
+ \ldots
\right.
\nonumber\\
& & \left. 
\qquad\ +\ 
 \lambda^n \ \Choose{12}{n}\ 
\varepsilon^{\alpha_1\alpha_2..\alpha_n\xi_1..\xi_{12-n}}\ 
\varepsilon_{\beta_1\beta_2..\beta _n\xi_1..\xi_{12-n}}\ 
\left(\ A\ \right)_{\alpha_1}^{\beta_1} 
\left(\ A\ \right)_{\alpha_2}^{\beta_2} 
\ldots \left(\ A\ \right)_{\alpha_n}^{\beta_n} 
\right.
\nonumber\\
& & \left.\qquad
\ \ldots\ \ +\ 
\lambda^{12} 
\varepsilon^{\alpha_1\alpha_2..\alpha_{12}}\ 
\varepsilon_{\beta_1\beta_2..\beta_{12}}\ 
\left(\ A\ \right)_{\alpha_1}^{\beta_1} 
\ldots \left(\ A\ \right)_{\alpha_{12}}^{\beta_{12}} 
\ \right]
\nonumber\\
&=&
\frac{1}{12!}\ \sum_{j=1}^{12}\ \Choose{n}{j} \ \lambda^j\  C_j(t)
\ \ \ ,
\label{eq:detA}
\end{eqnarray}
where in the last line we identify the matrix $A$ with $\Pi$.
Further,
\begin{eqnarray}
\det\left(1+\lambda A\right) & = & 
\exp\left({\rm Tr}\left[  \log\left[\ 1+\lambda A\right]\ \right]\ \right)
\ =\ 
\exp\left({\rm Tr}\left[ \sum_{p=1} {(-)^{p-1}\over p} \lambda^p A^p \right]\
\right)\
\nonumber\\
& = &
1\ +\ \lambda\ {\rm Tr}\left[\ A\ \right]
\ +\ 
{\lambda^2\over 2}\ \left(\ 
\left(  {\rm Tr}\left[\ A\ \right] \right)^2
\ -\ 
{\rm Tr}\left[\ A^2\ \right] 
\right)
\nonumber\\
& &\ +\ 
{\lambda^3\over 6}\ \left(\ 
2 {\rm Tr}\left[\ A^3\ \right] \ -\ 
3  {\rm Tr}\left[\ A\ \right]   {\rm Tr}\left[\ A^2\ \right] \ +\ 
 \left(\ {\rm Tr}\left[\ A\ \right] \right)^3\ \right)
\ +\ \ldots
\ .
\label{eq:detB}
\end{eqnarray}
Therefore, by equating terms of the same order in the expansion
parameter $\lambda$ in eq.~(\ref{eq:detA}) and eq.~(\ref{eq:detB}),
one can recover the $n$-$\pi^+$ correlation functions in
eq.~(\ref{eq:Cnepep}).  As an example, the contractions for the
$3$-$\pi^+$ system are
\begin{eqnarray}
C_3(t) & \propto & 
{\rm tr_{C,S}}\left[ \Pi \right]^3
\ -\  3\  {\rm tr_{C,S}}\left[ \Pi^2 \right] {\rm tr_{C,S}}\left[\Pi\right]
\ +\  2\  {\rm tr_{C,S}}\left[ \Pi^3 \right]
\ \ \ ,
\label{eq:threePiCorrelator}
\end{eqnarray}
where the traces, ${\rm tr_{C,S}}$, are over color and spin indices.
The three contributions in the correlator in
eq.~(\ref{eq:threePiCorrelator}) are shown in
fig.~\ref{fig:3+5pi_contractions}, (a), (b), and (c), respectively.
As it is the energy of states with maximal $z$-component of isospin
that are calculated in this work, disconnected contractions, such as
those in fig.~\ref{fig:3+5pi_contractions}(d), do not contribute to
the correlation functions that are computed.
The explicit form of the contractions for $n=1,\ldots 13$ are given in
Appendix \ref{sec:contr-part}.  Rewriting the contractions in terms of
traces over the $\Pi$-blocks greatly reduces the required number of
calculations, with the number of independent contributions to the
correlation function equal to the partition, $P(n)$, of $n$ objects.
An estimate of the number of operations that must be performed to
generate the correlator for $n$-mesons is $\sim n (12\times 13 -1) +
\sum_{j=1}^n P(j)$, which for large $n$ scales as $\sim {1\over
  2\sqrt{2}\pi \sqrt{n}} e^{\pi \sqrt{2 n/3}}$ using a classic result
of Hardy and Ramanujan~\cite{HardyRamanujan}.  While for $n=12$ there
are $\sim 2.3\times 10^{17}$ independent contractions that must be
performed, this can be accomplished with $\sim 2\times 10^3$
calculations to produce the $\sim 80$ terms contributing to the
contraction.  Since, in this work, each contraction is performed with
only a single quark propagator on each configuration, the
Pauli-exclusion principle requires that the $n \ge 13$ identical meson
contractions vanish identically, e.g.  $C_{13}(t)=0\ \forall\ t$,
implying the $\lambda^{13}$ and higher terms in the expansion of
eq.~(\ref{eq:detB}) vanish. Written in terms of contractions of
propagators in flavor and color space, the $n=13$ case of
eq.~(\ref{eq:detB}), represents a generalized Cayley-Hamilton identity
satisfied by all matrices of size less than $13\times 13$. To perform
calculations on systems containing more than twelve pions, additional
propagators will be required.

\subsection{High-precision implementation}
\label{sec:high-prec-impl}

\begin{figure}[!ht]
  \centering
  \includegraphics[width=0.7\columnwidth]{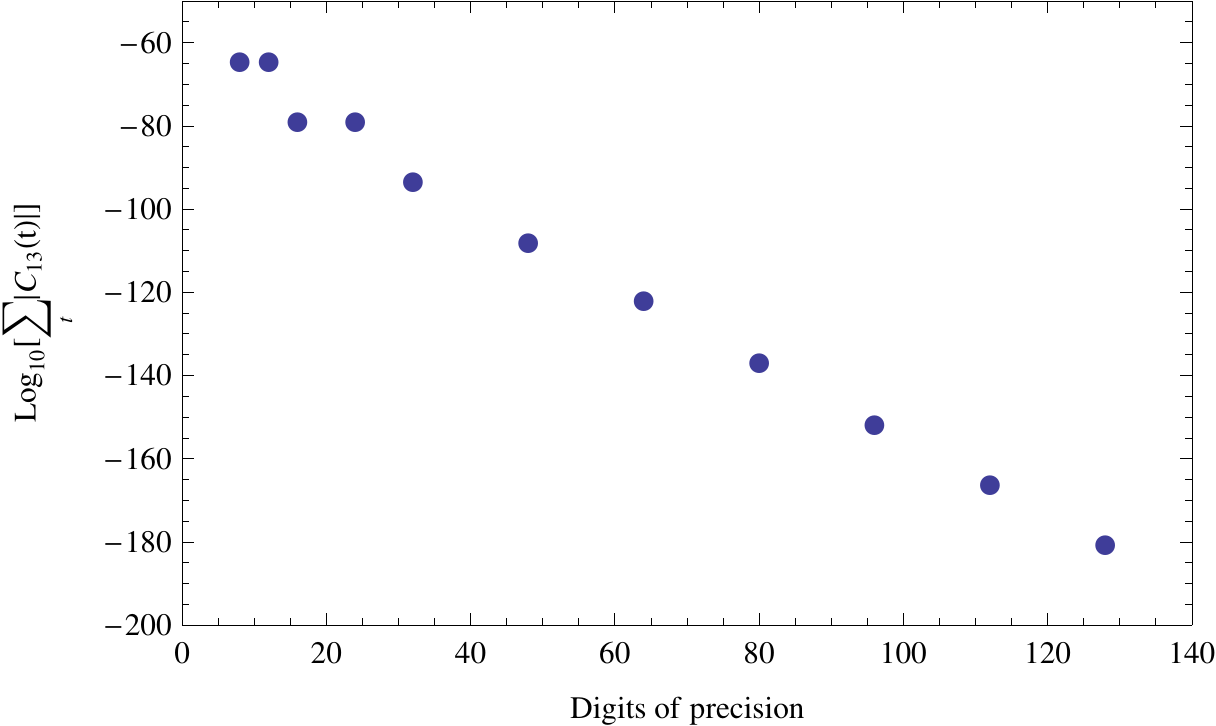}
  \caption{The $n=13$ correlation function as a function of the
    precision used to perform the calculation. The vertical axis is
    the logarithm of the sum over time-slices of the absolute value of
    the $n=13$ correlation function on a representative gauge field
    configuration.}
  \label{fig:n13}
\end{figure}
\noindent
In order to calculate the $n$-$\pi^+$ correlation functions,
particularly for $n \ge 8$, it is necessary to use a numerical
representation with precision greater than that of standard 64-bit
machine precision.\footnote{ The failure of double precision
  operations for these correlation functions is explored in detail in
  Appendix~\ref{sec:numer-prec-n}.}  This need arises because of the
large products of propagators that must be computed and is not
particular to the contractions studied here.  In particular, the
numerical issues impacting calculations of multi-pion systems that we
have found in this work will also impact calculations of multi-nucleon
systems.

Our implementation of these contractions uses arbitrary precision
arithmetic based on the ${\tt ARPREC}$ library \cite{ARPREC} which was
extended for the particular operations needed here, matrix
multiplications and traces. For the correlators studied here 64
decimal digit precision (approximately octupule precision) in internal
operations is sufficient to give results accurate to sixteen digits.
The additional overhead of using this numerical representation causes
the high precision contraction code to run $\sim$10--50 times slower
than a double-precision version but is only marginally dependent on
the precision used\footnote{Checks of our {\tt c++} contractions have
  also been performed using {\it Mathematica 6.0} (time costs prevent
  us using this on a large scale).}.  For the $n=13$ correlation
function it is instructive to look at the dependence of the resulting
correlator on the precision used in the computations.  Since the
correlation function must vanish identically for any input propagator,
it is a very stringent test of the codes used herein.  In
fig.~\ref{fig:n13}, the logarithm of the sum over time-slices of the
absolute value of the correlation function as a function of the digits
of precision used to perform the contractions on a representative
configuration is shown. From extrapolating the results shown here, we
conclude that the correlator is indeed identically zero.

\subsection{Analysis}
\label{sec:fitting-procedures}

\noindent
The correlation functions from which we extract the ground-state
energy of the $n$-$\pi^+$ system are given in eq.~(\ref{eq:Cnfun}),
and, on a lattice with infinite extent in the time direction, behave
as
\begin{eqnarray}
C_n(t)&\stackrel{t\to\infty}{\longrightarrow} & {\cal A}_0^{(n)}\ e^{- E_n\ t}
\label{eq:cnlarget}
\end{eqnarray}
at large times.  It is the difference between this energy, $E_n$ and
$n$ times the $\pi^+$ rest mass that is equated to the energy
difference given in eq.~(\ref{eq:energyshift}), and which is extracted
from the ratio of correlation functions
\begin{eqnarray}
G_n(t) 
&  = & { C_n(t) \over \left[\ C_1 (t)\ \right]^n }
\ \stackrel{t\to\infty}{\longrightarrow}\ {\cal B}_0^{(n)}\ e^{- \Delta E_n\ t}
\ \ \ ,
\label{eq:Gnlarget}
\end{eqnarray}
where $\Delta E_n$ is that of eq.(\ref{eq:energyshift}).  While there
are a number of ways to extract the energy difference from the
correlation function, perhaps the most visually pleasing one is to
construct the effective energy difference function, defined to be
\begin{eqnarray}
\Delta E^{\rm eff.}_n (t) & = & 
\log\left( { G_n(t)\over G_n(t+1)}\right) 
\ \stackrel{t\to\infty}{\longrightarrow}\ \Delta E_n
\ \ \ .
\label{eq:Eeff}
\end{eqnarray}
In the limit of an infinite number of measurements, this function
would tend to a constant equal to the ground-state energy splitting.
Of course, for any real calculation, both the number of gauge fields
and the number of propagators per gauge field are finite, and as such
the object $\Delta E_n^{\rm eff.} (t) $ consists of a central value
and an associated uncertainty at each time-slice, $t$.  Further, the
temporal extent of the lattice is finite, giving rise to both forward
and backward propagating hadrons. As such, $\Delta E_n^{\rm eff.}  (t)
$ is constant (up to statistical fluctuations) only over a finite
number of time-slices, in a region between where the forward
propagating excited states have died-out sufficiently, and where the
backward propagating states have not yet become significant.  In the
current situation where Dirichlet boundary conditions are imposed, the
behavior of the correlator is modified by the ``reflection'' of
forward and backward propagating states from the boundaries.  As these
reflections are poorly understood, the region close to the boundary is
omitted in our analysis.

In our lattice calculations, multiple propagators and correlation
functions are computed on each gauge configuration.  These propagators
and correlation functions are not statistically independent unless the
sources are separated by many correlation lengths.  To account for
this, all the correlation functions for a fixed $n$ computed on a
given configuration are averaged (blocked) into one correlation
function.  The configurations are found to be statistically
independent\footnote{This is tested by averaging over sets of
  neighboring configurations and performing analysis on the resulting
  blocked ensemble. For block sizes of 1, 4 and 12, no noticeable
  difference is seen.}, and the blocked correlators on each
configuration form the basis of our statistical analysis.

In order to extract the energy difference $\Delta E_n$ from $\Delta
E_n^{\rm eff.}  (t) $, a fitting interval must be selected. This
interval is chosen to be entirely contained in the region where the
$\Delta E_n^{\rm eff.}(t) $ is consistent with a constant.  Once the
fitting interval has been selected a correlated $\chi^2$ minimization
is performed to extract the parameter $\Delta E_n$, defined in
eq.~(\ref{eq:Gnlarget}).  The covariance matrix that determines the
correlated weightings of each of the values of $\Delta E_n^{\rm eff.}
(t) $ on any given time-slice is generated using a single-subtraction
Jackknife procedure.\footnote{As a check, we also performed a separate
  analysis using bootstrap re-sampling. The resulting energies and
  parameters were consistent, and, for simplicity, we focus on a
  single analysis in the main discussion.}  The central value of
$\Delta E_n$ is the value that minimizes the correlated $\chi^2$, and
the standard statistical uncertainty is determined by the values of
$\Delta E_n$ for which $\chi^2\rightarrow\chi^2+1$.  The fitting
systematic uncertainty associated with the fitting procedure is
determined by varying each end of the fitting range by $-2 \le \Delta
t \le +2$, and refitting the energy-splitting.

To study the scattering length and three-body parameter, \sheep, using
eqs.~(\ref{eq:abarisolation}) and (\ref{eq:eta3barbarisolation}), the
appropriate ratios of the $C_n(t)$ correlators are used to define
effective scattering length functions for each $n$ (LO, NLO, N$^2$LO)
or each pair \{$n$,$m$\} (N$^3$LO and N$^4$LO) and effective \sheep\ 
functions for each $n$. We then analyze these in the same manner as
the energy differences above. This leads to multiple determinations of
$\apipi$ and \sheep. The effective functions defined by $\zeta_{6,7}$,
eqs.~(\ref{eq:vanishesL6}) and (\ref{eq:vanishesL7}), are studied
similarly.

To make use of the full data set, we also perform a simultaneous,
correlated fit of $\apipi$ and \sheep\ to the effective masses for
$n=2,3,\ldots,N_{\rm max}$ for $N_{\rm max}=3,\ldots,12$.  In order to
do this, fitting ranges, $ t_{\rm min}^{(n)} \le t \le t_{\rm
  max}^{(n)}$, are chosen for each $n$ (as above) and the data is
assembled into a vector $V = \{ \Delta E_2^{\rm eff.}  (t_{\rm
  min}^{(2)}), \ldots \Delta E_2^{\rm eff.}  (t_{\rm max}^{(2)}),
\Delta E_3^{\rm eff.} (t_{\rm min}^{(3)}), \ldots \Delta E_3^{\rm
  eff.} (t_{\rm max}^{(3)}), \ldots\Delta E_{12}^{\rm eff.} (t_{\rm
  min}^{(12)}), \ldots \Delta E_{12}^{\rm eff.}  (t_{\rm max}^{(12)})
\}$.  A correlated $\chi^2$ minimization is performed to extract the
parameters $\apipi$ and \sheep\ via the fit vector $U=\{\Delta E_2
,\ldots , \Delta E_2,\Delta E_3 ,\ldots , \Delta E_3 ,\ldots,\Delta
E_{12} ,\ldots , \Delta E_{12} \}$, where $\Delta E_n$ is given in
terms of $\apipi$ and \sheep\ using eq.~(\ref{eq:energyshift}), and
where the covariance matrix that determines the correlated weightings
of each contribution is generated using the Jackknife procedure.  The
standard (statistical) uncertainties on $\apipi$ and \sheep\ are
determined from the maximum and minimum values of each parameter
dictated by the uncertainty-ellipse corresponding to their values for
which $\chi^2\rightarrow\chi^2+1$.  The systematic uncertainty
associated with the fitting procedure is determined by repeatedly and
randomly varying each end of the fitting range for each correlator by
$-2 \le \Delta t \le +2$, refitting the parameters $\apipi$ and
$\overline{\overline{\eta}}_3^L$, determining the complete range of
values for each parameter associated with
$\chi^2\rightarrow\chi^2+1$.\footnote{An alternative systematic
  procedure of repeatedly and randomly choosing triplets of
  time-slices in each fit range $\pm1$ time-slice and refitting is
  also used, giving qualitatively similar results.}  The systematic
and statistical uncertainties are combined in quadrature in this work.

\section{Lattice QCD results}
\label{sec:correlation-energies}

\noindent
Using the techniques discussed in the previous section, we now turn to
analysis of the results of the lattice calculations.  Our main aim is
to extract the parameters $\apipi$ and \sheep\ which we can do in a
number of ways, either by forming particular combinations of energies,
eqs.~(\ref{eq:abarisolation}) and (\ref{eq:eta3barbarisolation}), or
by a coupled analysis accounting for correlations among different $n$.
The different methods give consistent results but we find that the
most precise extraction is achieved using the latter method and
consequently our final results are generated from this technique.
Before we present these results, we first detail the simpler analysis
using combinations of energies. As there are a large number of
correlation functions that we study in this work, in some intermediate
stages, we only display results for a single quark mass corresponding
to $m_\pi = 291$~MeV (in terms of uncertainties, this ensemble is
neither the best nor the worst).

\subsection{Multi-pion energies and energy differences}
\label{sec:multipion-energies}

\begin{figure}[!t]
  \centering
  \includegraphics[width=0.90\columnwidth]{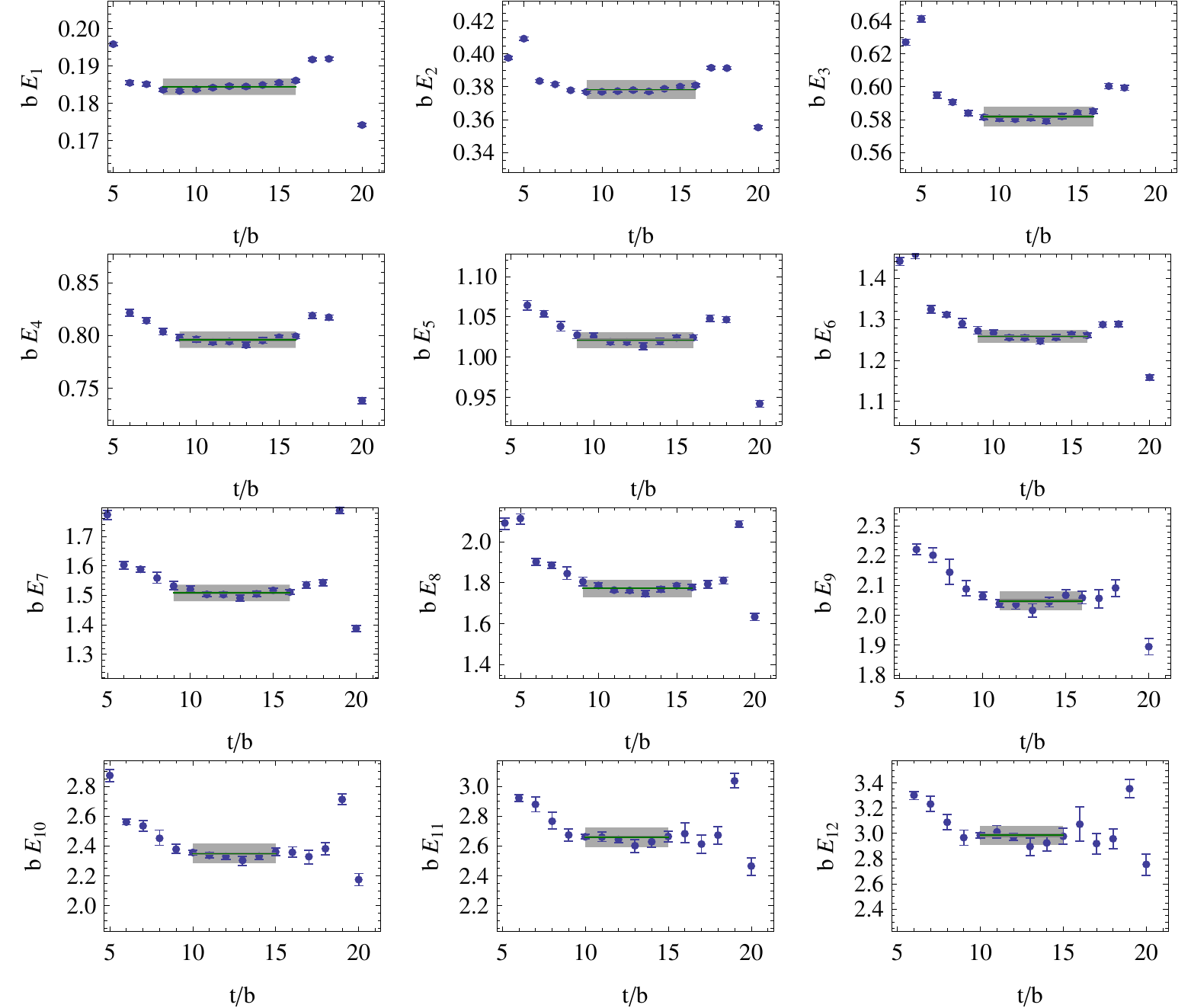}
  \caption{Effective energy plots for the $n=1,\ldots,12$ pion correlations as
    a function of time for the ensemble with $m_\pi=291$~MeV. The
    solid line and shaded region show the fitted energy and the
    systematic and statistical uncertainties combined in quadrature.}
  \label{fig:energy007pi}
\end{figure}

\begin{figure}[!t]
  \centering
  \includegraphics[width=0.45\columnwidth]{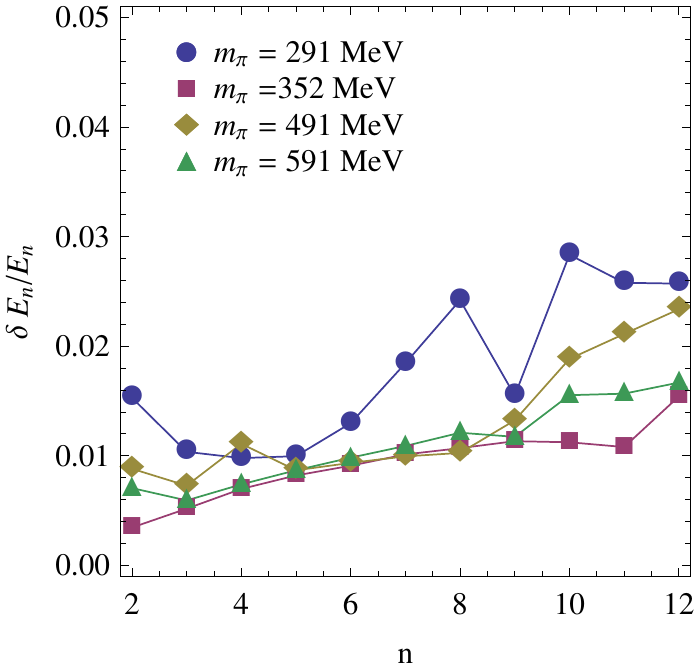}\qquad
  \includegraphics[width=0.45\columnwidth]{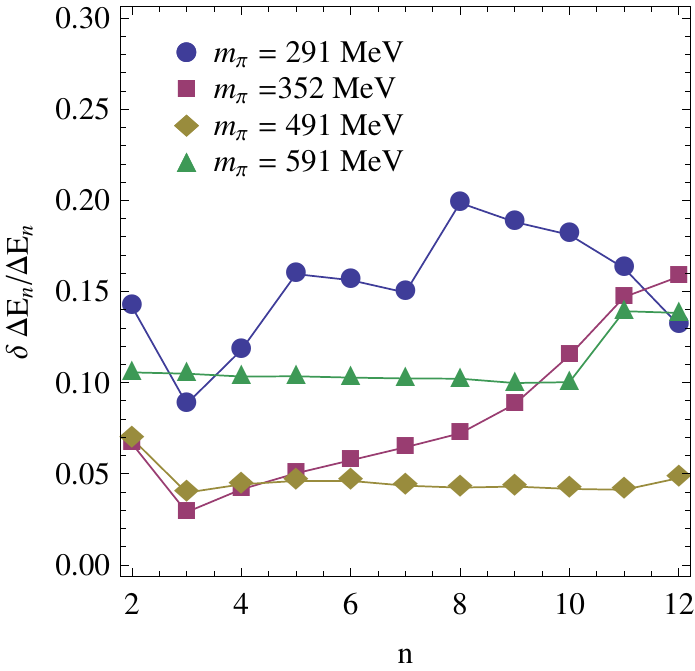}
  \caption{Relative uncertainties in extractions of energies (left) and
    energy differences (right) as a function of the number of
    $\pi^+$s.  Results are shown for all quark masses.}
  \label{fig:relerrenergy}
\end{figure}

\noindent
{\it A priori}, it may seem surprising that the correlation function
of twelve $\pi^+$'s can be calculated at all.  On the ensemble
associated with the lightest pion mass, $m_\pi = 291~{\rm MeV}$, the
$12$-$\pi^+$ state has an energy of $E_{12}\sim 3.5~{\rm GeV}$, while
on the ensemble with $m_\pi = 591~{\rm MeV}$, the $12$-$\pi^+$ state
has an energy of $E_{12}\sim 7.1~{\rm GeV}$. It is not immediately
obvious that such a rapidly diminishing exponential can be cleanly
measured but in these systems we find that it is possible.\footnote{In
  purely pionic systems there is no exponential degradation of the
  signal-to-noise ratio, unlike in most other hadronic
  systems~\cite{Lepage:1989hd}.}  As an example, for $m_\pi = 291~{\rm
  MeV}$, the effective energies (in lattice units) for systems with
$n=1,2,.. ,12$ $\pi^+$'s are shown in fig.~\ref{fig:energy007pi}.
Well-defined plateaus in the effective energy plots are seen for all
systems, with the relative statistical uncertainty in the data almost
constant as a function of $n$. As can be seen from the fits to the
energies (statistical and systematic uncertainties are shown in
quadrature), the precision with which the energy can be extracted is
high, typically $<2$\%. The total relative uncertainties on the energy
of the $n$-$\pi^+$ systems are shown for all data sets in
fig.~\ref{fig:relerrenergy}, with only a slight dependence of $n$
apparent.
An additional point is that it was not obviously the case that the
Gaussian-smeared source for the light-quark, that is suitable for the
single pion ground-state, would have sufficient overlap onto the
multi-pion ground-state to produce useful correlation functions.
However, it is clear that it did.

\begin{figure}[!t]
  \centering
  \includegraphics[width=0.90\columnwidth]{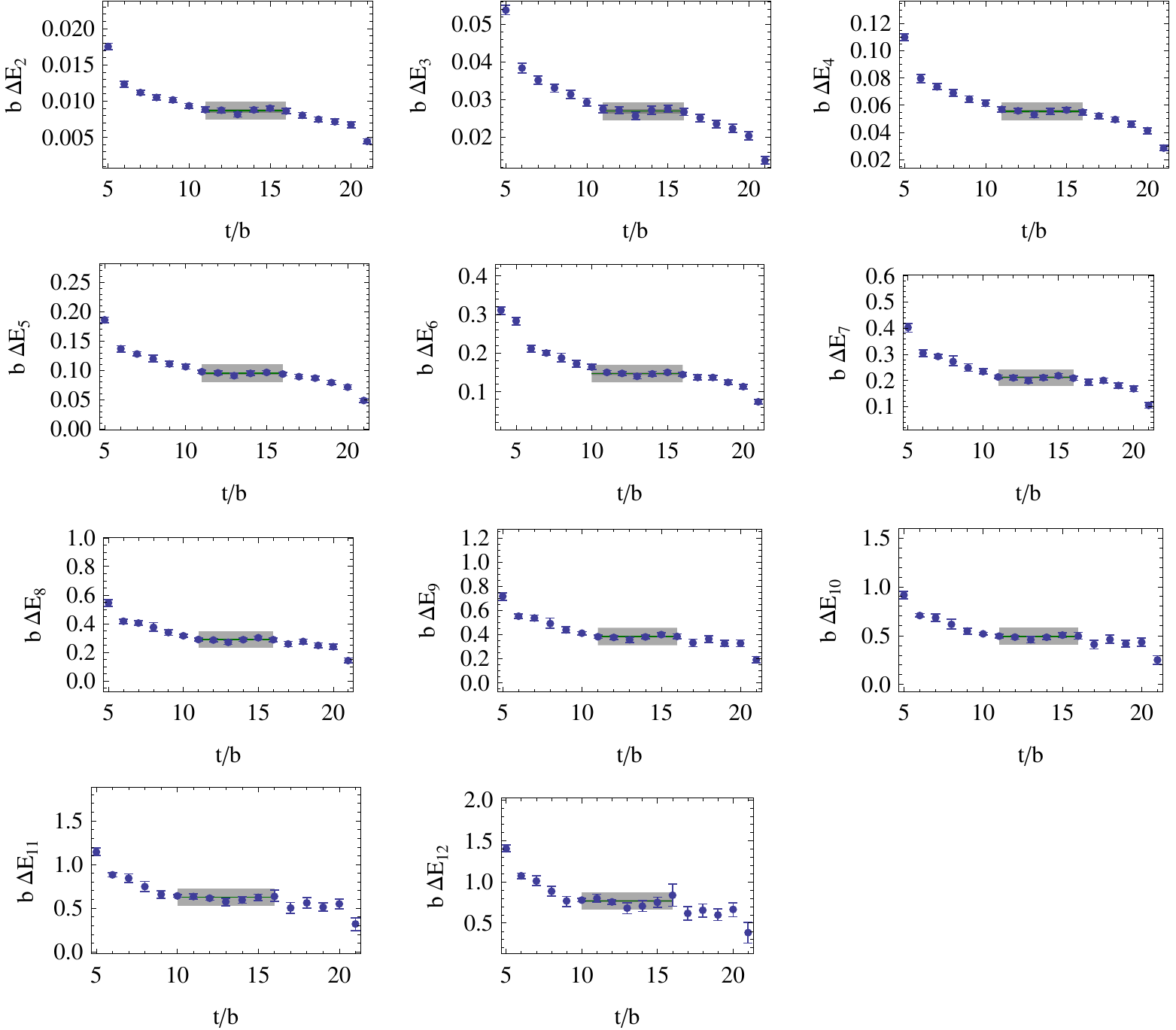}
  \caption{Effective energy difference plots for $n=2,\ldots,12$ pion
    correlations as a function of time for the ensemble at $m_\pi =
    291$~MeV. The solid line and shaded region show our fitted energy
    and the systematic and statistical uncertainties combined in
    quadrature.}
  \label{fig:energydiff007}
\end{figure}

The energy differences which enter into eq.~(\ref{eq:energyshift}) and
subsequent results can also be extracted cleanly, although with
somewhat less precision than the individual energies. The effective
energy difference plots, along with our fits to the energy differences
are shown in fig.~\ref{fig:energydiff007} (again for the
$m_\pi=291$~MeV ensemble) while the relative uncertainties in our
extractions are given in fig.~\ref{fig:relerrenergy}. All effective
energy splitting plots show behavior that is consistent with a single
exponential (within statistical uncertainties) for a number of time
slices. As discussed above, the region above $t/b\sim16$ is
contaminated by reflections from the Dirichlet boundary at $t/b=22$
and is discarded in our analysis.

\subsection{$\pi^+\pi^+$ scattering length}
\label{sec:pipi-scatt-length}

\noindent
Since the $\pi^+\pi^+$ system is free from three- and higher-body
hadronic interactions, it is the ideal place to extract the two body
parameter, $\apipi$. As is well known, this can be done without
resorting to an expansion in $\apipi/L$ using the eigenvalue equation
in eq.~(\ref{eq:energies}).  We refer to $\apipi$ extracted in this
way as the L\"uscher result and it forms a benchmark for extractions
in the $n>2$ systems.  Eq.~(\ref{eq:energyshift}) also allows us to
extract $\apipi$ in a number of ways. At orders $L^{-3}, L^{-4},
L^{-5}$ (LO, NLO and N$^2$LO, respectively), each energy difference,
$\Delta E_n$ for $n=2,\ldots,12$, leads to a separate extraction of
$\apipi$. Finally eq.~(\ref{eq:abarisolation}), allows us to extend
these extractions to N$^3$LO and N$^4$LO in the $\apipi/L$ expansion
by combining the $n$- and $m$- body energy differences to eliminate
three-body interactions. Choosing $3\leq m<n\leq12$, allows for 45
separate extractions.

\begin{figure}[!t]
  \centering
  \includegraphics[width=0.90\columnwidth]{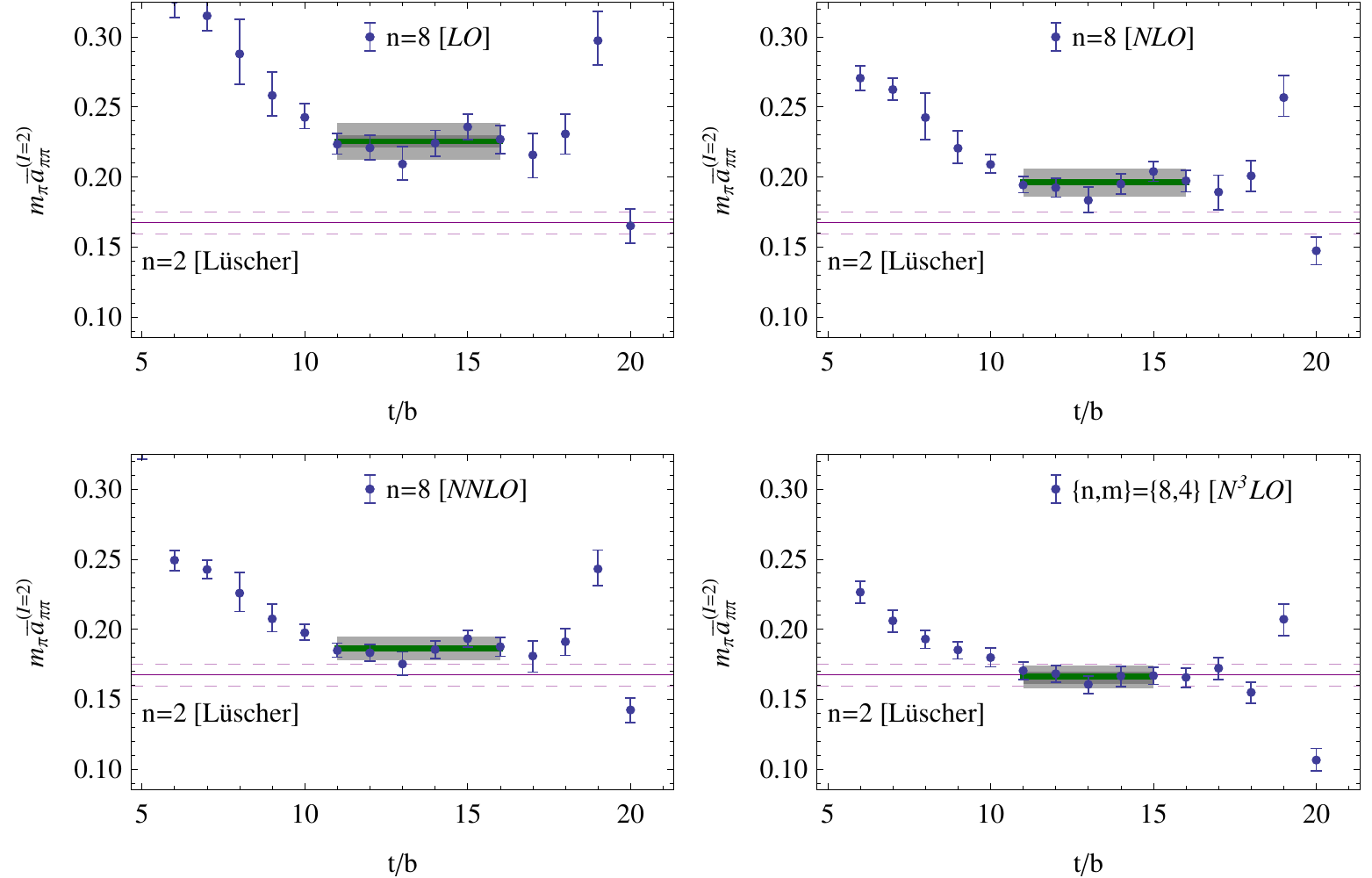}
  \caption{Effective $m_\pi \apipi$ plots for the LO (top-left), NLO
    (top-right) and N$^2$LO (bottom-left) extractions using the $n=8$
    energy and the N$^3$LO effective $m_\pi \apipi$ extracted from
    $\{n,m\}=\{8,4\}$ (bottom-right) from
    eq.~(\protect\ref{eq:abarisolation}).  In each case the
    $m_\pi=291$~MeV ensemble is used and the scale in each plot is
    identical. The horizontal band corresponds to the extraction using
    the exact eigenvalue method for $n=2$. }
  \label{fig:effabar}
\end{figure}
\begin{figure}[!th]
  \centering
  \includegraphics[width=0.45\columnwidth]{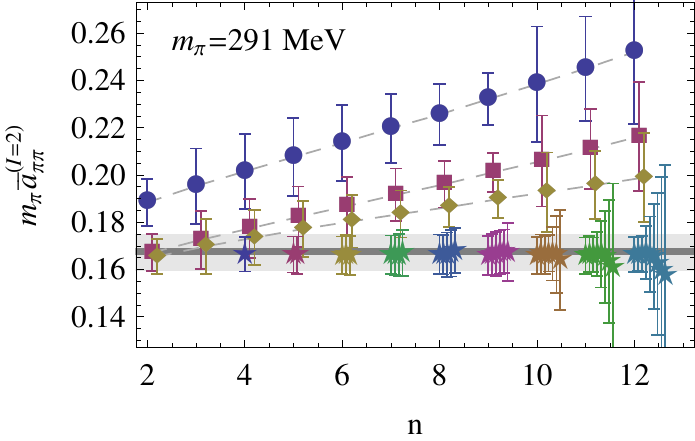}\qquad
  \includegraphics[width=0.45\columnwidth]{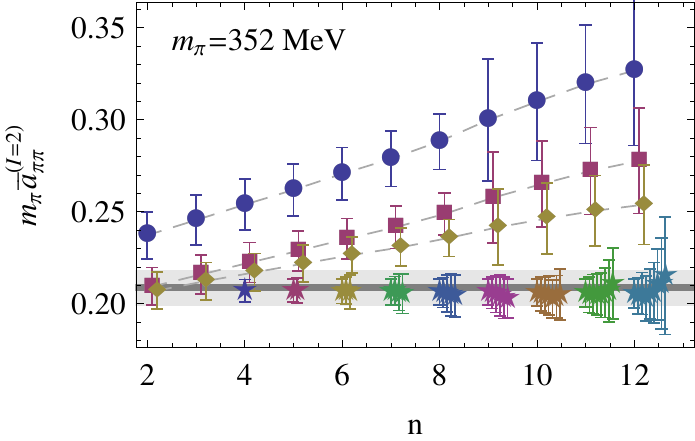}
  \\

  \includegraphics[width=0.45\columnwidth]{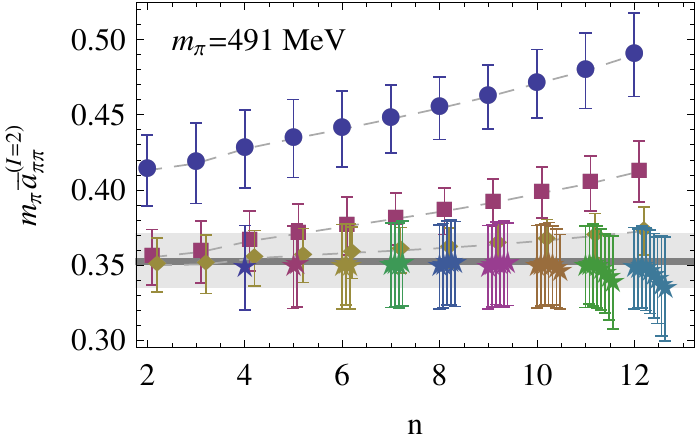}\qquad
  \includegraphics[width=0.45\columnwidth]{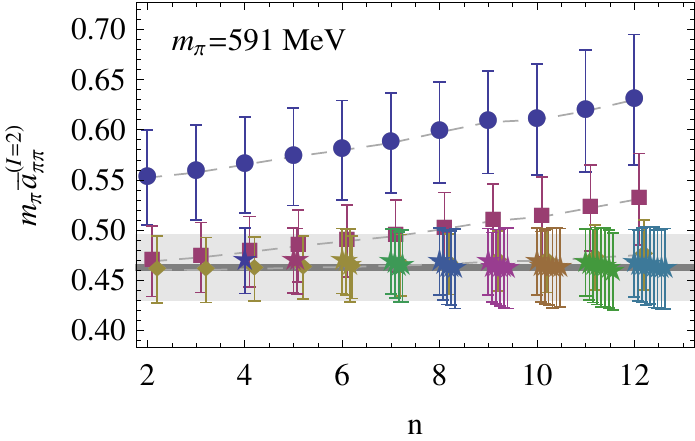}
  \caption{Extractions of $m_{\pi} \apipi$ 
    for each ensemble.  The blue circles, mauve squares and brown
    diamonds correspond to the extractions at ${\cal O}(L^{-3})$,
    ${\cal O}(L^{-4})$, and ${\cal O}(L^{-5})$ in the $1/L$ expansion
    given in eq.~(\protect\ref{eq:energyshift}).  The stars correspond
    to the ${\cal O}(L^{-6})$ extractions using
    eq.~(\protect\ref{eq:abarisolation}), which requires two different
    energy shifts. At any given $n$, we have shown various
    combinations of $m<n$.  Finally the solid line and the shaded
    region correspond to the extraction of $m_{\pi} \apipi$ using the
    exact eigenvalue method of L\"uscher from the $n=2$ data. In all
    cases, statistical and systematic uncertainties have been combined
    in quadrature.  The upper-left, upper-right, lower-left and
    lower-right panels corresponds to $m_\pi = 291, 352, 491, 591
    ~{\rm MeV}$, respectively.  }
  \label{fig:mpia_summary}
\end{figure}

As a representative example, fig.~\ref{fig:effabar} shows the LO, NLO
and N$^2$LO effective $\apipi$ plots\footnote{Given the accuracy with
  which $m_\pi$ and $f_\pi$ have been extracted on these ensembles,
  the uncertainties in $m_\pi \apipi$ and $m_\pi f_\pi^4$\sheep are
  dominated by the uncertainties in $\apipi$ and \sheep,
  respectively.}  for $n=8$, and the N$^3$LO effective $\apipi$ plots
for $\{n,m\}=\{8,4\}$ from eq.~(\ref{eq:abarisolation}), for the
$m_\pi = 291$~MeV ensemble.  A summary of all the extractions of
$\apipi$ is given in fig.~\ref{fig:mpia_summary} (the N$^4$LO
extractions are entirely consistent with those at N$^3$LO in all cases
and are omitted).  For the $n=2$ data, it is clear that NLO and higher
extractions, fig.~\ref{fig:mpia_summary}, yields the same $\apipi$ as
the exact eigenvalue method of L\"uscher. However, for the multi-pion
systems, an $n$-dependent systematic deviation from the exact
eigenvalue method of L\"uscher is found at LO, NLO and N$^2$LO.  This
is particularly clear for the lighter mass ensembles. In contrast, the
extractions of $\apipi$ in which the $\pi^+\pi^+\pi^+$-interaction (or
at least, a term that behaves as $\Choose{n}{3}$) is eliminated
(N$^3$LO and N$^4$LO) are in close agreement with the $n=2$ exact
eigenvalue result for all $n$. From this alone we conclude that the
calculation suggests the presence of a significant
$\pi^+\pi^+\pi^+$-interaction.

\begin{figure}[!t]
  \centering
  \includegraphics[width=0.90\columnwidth]{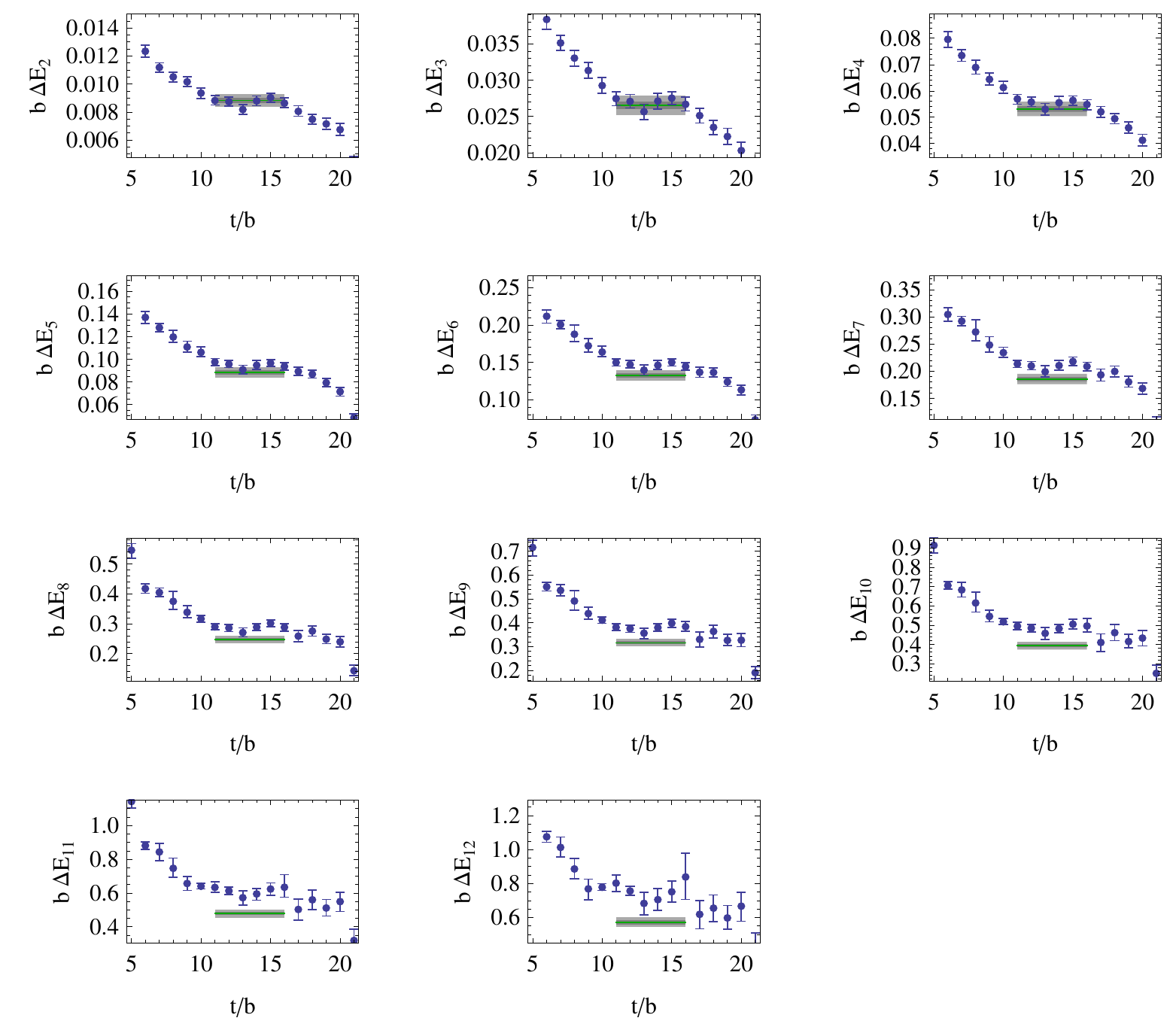}
  \caption{
    The effective energy splitting plots for the $m_\pi = 291$~MeV
    ensemble.  The solid lines correspond to the energy differences of
    eq.~(\ref{eq:energyshift}) using the value of $\apipi$ from the
    $\pi^+\pi^+$ energy-splitting (using the exact eigenvalue method)
    and setting $\overline{\overline{\eta}}^L_3=0$. The shaded region
    shows the statistical and systematic uncertainties combined in
    quadrature. }
  \label{fig:energydiffsandfitscompare}
\end{figure}

We can further demonstrate the need for a
$\pi^+\pi^+\pi^+$-interaction by using eq.~(\ref{eq:energyshift}) to
compute the energy shifts at ${\cal O}(L^{-7})$ using the value of
$\apipi$ extracted from the $\pi^+\pi^+$-system using the exact
eigenvalue method and setting \sheep$=0$. These can then be compared
to the calculated effective energy differences as shown in
fig.~\ref{fig:energydiffsandfitscompare} for the $m_\pi = 291$~MeV
ensemble.  The deviations between the predictions and the effective
energy splittings are significant and grow with increasing $n$.  Over
the same sets of time-slices as used in the fits to the two-particle
energy difference, the $\chi^2/d.o.f.$ of such an ansatz is 8.62 and
therefore \sheep$=0$ very poorly describes the results of the
calculation.

\subsection{Three body interaction}
\label{sec:three-body-inter}

\begin{figure}[!t]
  \centering
  \includegraphics[width=0.90\columnwidth]{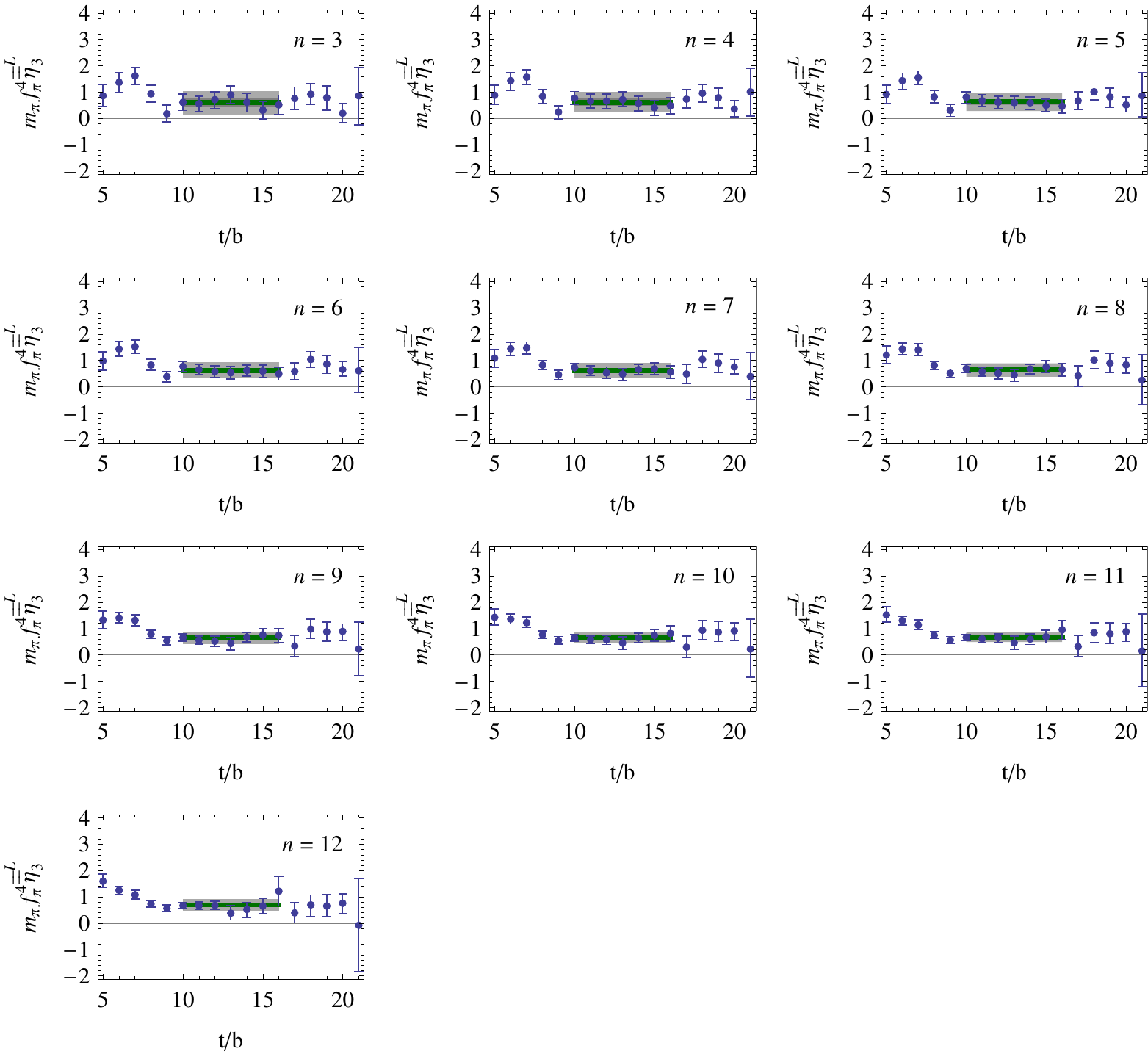}
  \caption{The effective \sheep\ plots for
    the $m_\pi = 291$~MeV ensemble using
    eq.~(\protect\ref{eq:eta3barbarisolation}) at ${\cal O}(L^{-7})$.
    Statistical and systematic uncertainties are added in quadrature
    (shaded band).}
  \label{fig:eta3bar_summary_010}
\end{figure}

\begin{figure}[!t]
  \centering
  \includegraphics[width=0.45\columnwidth]{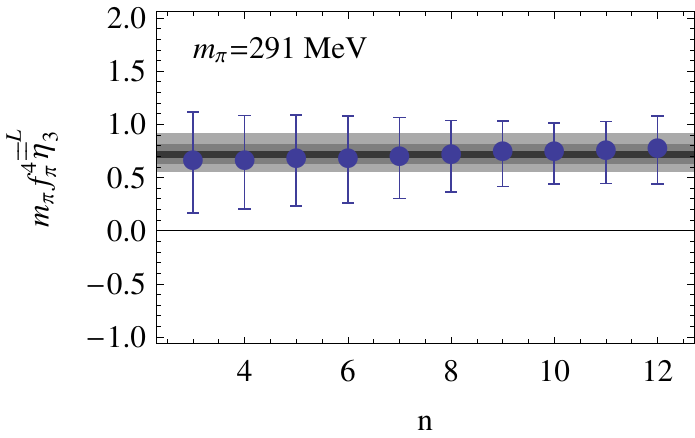}\qquad
  \includegraphics[width=0.45\columnwidth]{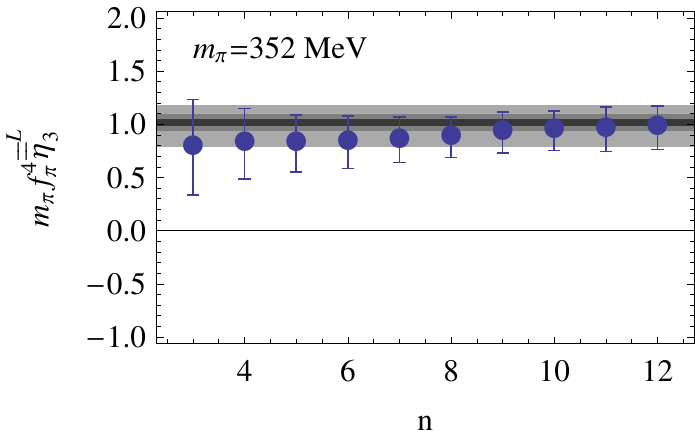}
  \\

  \includegraphics[width=0.45\columnwidth]{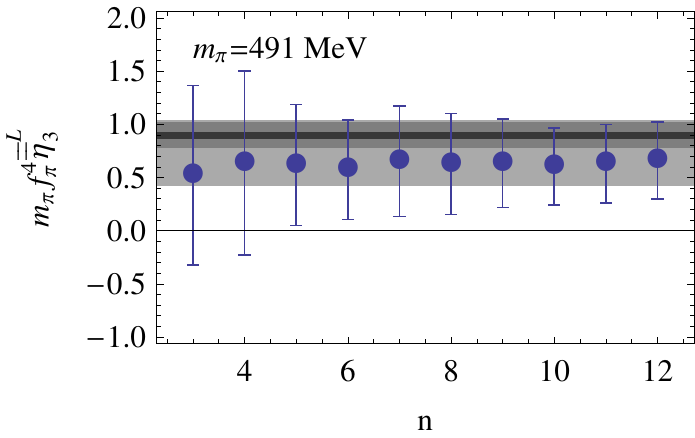}\qquad
  \includegraphics[width=0.45\columnwidth]{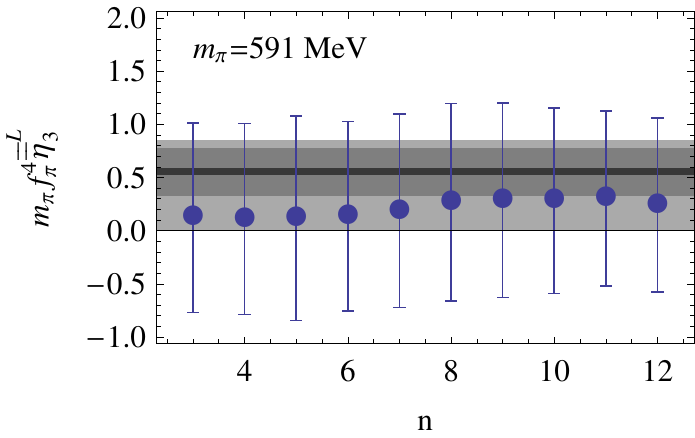}
  \caption{The value extracted for \sheep\ using
    eq.~(\protect\ref{eq:eta3barbarisolation}) at ${\cal O}(L^{-7})$
    as a function of $n$ for each ensemble used in the calculation.
    The horizontal line in each plot corresponds to the value
    extracted using all the data in the $n$-correlated fit of
    Section~\protect\ref{sec:n-corr-analys}. The inner and outer
    shaded bands correspond to the statistical uncertainty and the
    statistical and systematic uncertainties combined in quadrature,
    respectively.  }
  \label{fig:eta3bar_summary}
\end{figure}

\noindent 
The $\pi^+\pi^+\pi^+$-interaction can be explicitly constructed using
eq.~(\ref{eq:eta3barbarisolation}) for $n=3,\ldots,12$.  As an
example, the effective \sheep\ plots for the $m_\pi = 291$~MeV
ensemble are shown in fig.~\ref{fig:eta3bar_summary_010}.  A clear
plateau inconsistent with zero is seen in most cases.
Fig.~\ref{fig:eta3bar_summary} shows the $n$ dependence of the
extracted value of \sheep\ for each quark mass.  In general, the
combined systematic and statistical uncertainty of the extractions
decreases with increasing number of $\pi^+$'s. This is not surprising
given the combinatoric factors that appear in the expression for the
energy shift.

\subsection{Convergence: $\zeta_{6,7}$}
\label{sec:conv-zeta_7-7}

\noindent
Before presenting the $n$-correlated analysis we briefly turn to the
quantities $\zeta_{6,7}$ defined in eqs.~(\ref{eq:vanishesL6}) and
(\ref{eq:vanishesL7}). A plateau in the corresponding
effective-$\zeta_{6,7}$ plots for a particular pair $\{n,m\}$ at a
value inconsistent with zero would signal the breakdown of the
large-volume expansion in eq.~(\ref{eq:energyshift}) that is central
to our analysis.  In all cases, no such breakdown is seen.  However
for increasing $n$ and $m$, the uncertainties increase. For example,
for the systems with $\{n,m\}=\{11,12\}$ at $m_\pi = 291$~MeV, these
quantities are found to be $\zeta_{6}=\zeta_{7}=0.0(3)$.

\subsection{$n$-correlated analysis}
\label{sec:n-corr-analys}

\begin{figure}[!t]
  \centering
  \includegraphics[width=0.90\columnwidth]{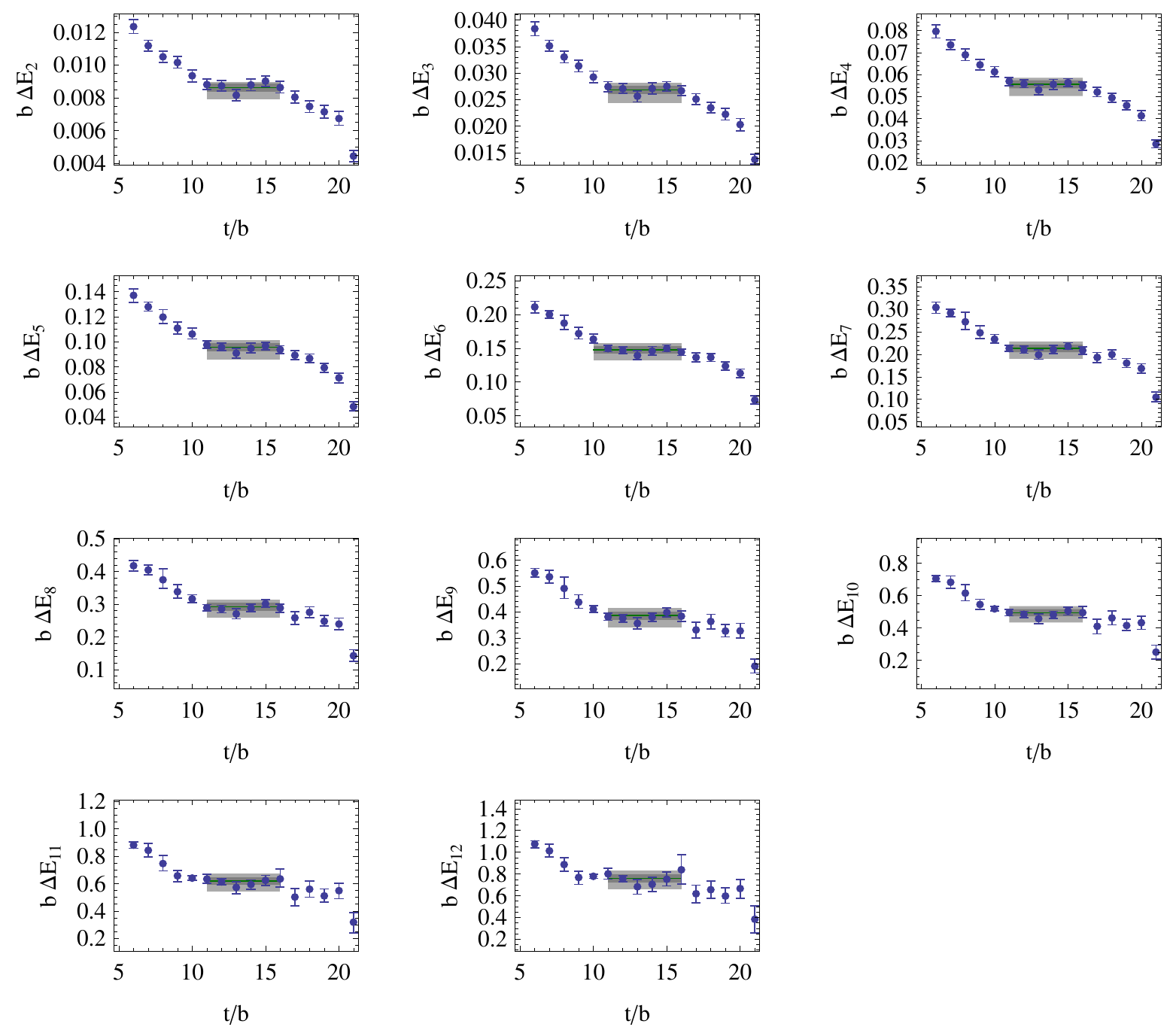}
  \caption{$n$-correlated fits to the $n=2,\ldots,12$ energy
    differences for the $m_\pi = 291$~MeV ensemble. All energy
    differences, $\Delta E_{\rm eff.}^{(n)}(t)$ for $n=2,\ldots,12$,
    are used.  Statistical, and statistical plus systematic (added in
    quadrature) uncertainties are shown as the inner and outer shaded
    regions, respectively.}
  \label{fig:corr007}
\end{figure}

\begin{figure}[!t]
  \centering
  \includegraphics[width=0.90\columnwidth]{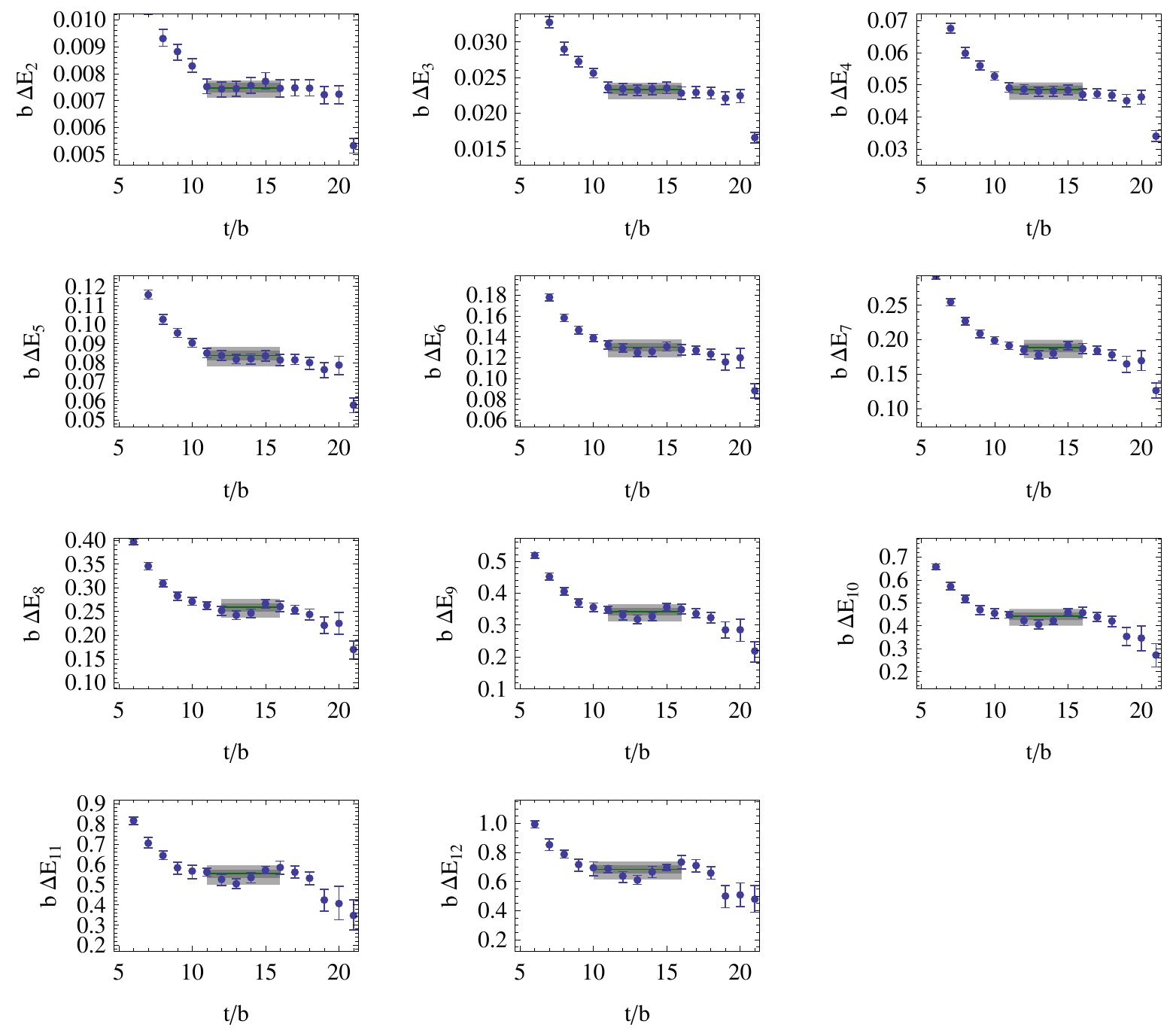}
  \caption{$n$-correlated fits to the $n=2,\ldots,12$ energy
    differences for the $m_\pi = 352$~MeV ensemble.}
  \label{fig:corr010}
\end{figure}

\begin{figure}[!t]
  \centering
  \includegraphics[width=0.90\columnwidth]{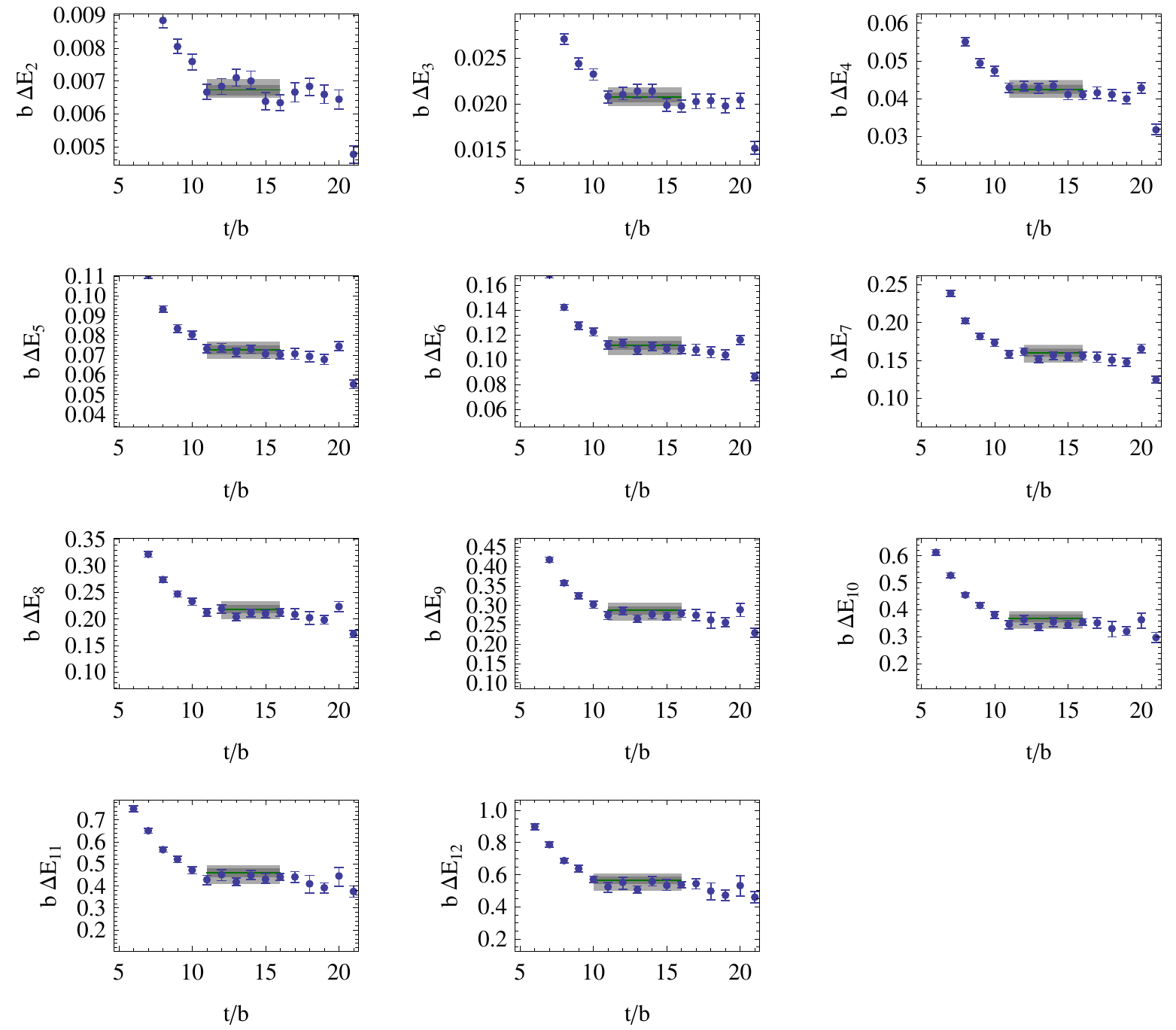}
  \caption{$n$-correlated fits to the $n=2,\ldots,12$ energy
    differences for the $m_\pi = 491$~MeV ensemble.}
  \label{fig:corr020}
\end{figure}

\begin{figure}[!t]
  \centering
  \includegraphics[width=0.90\columnwidth]{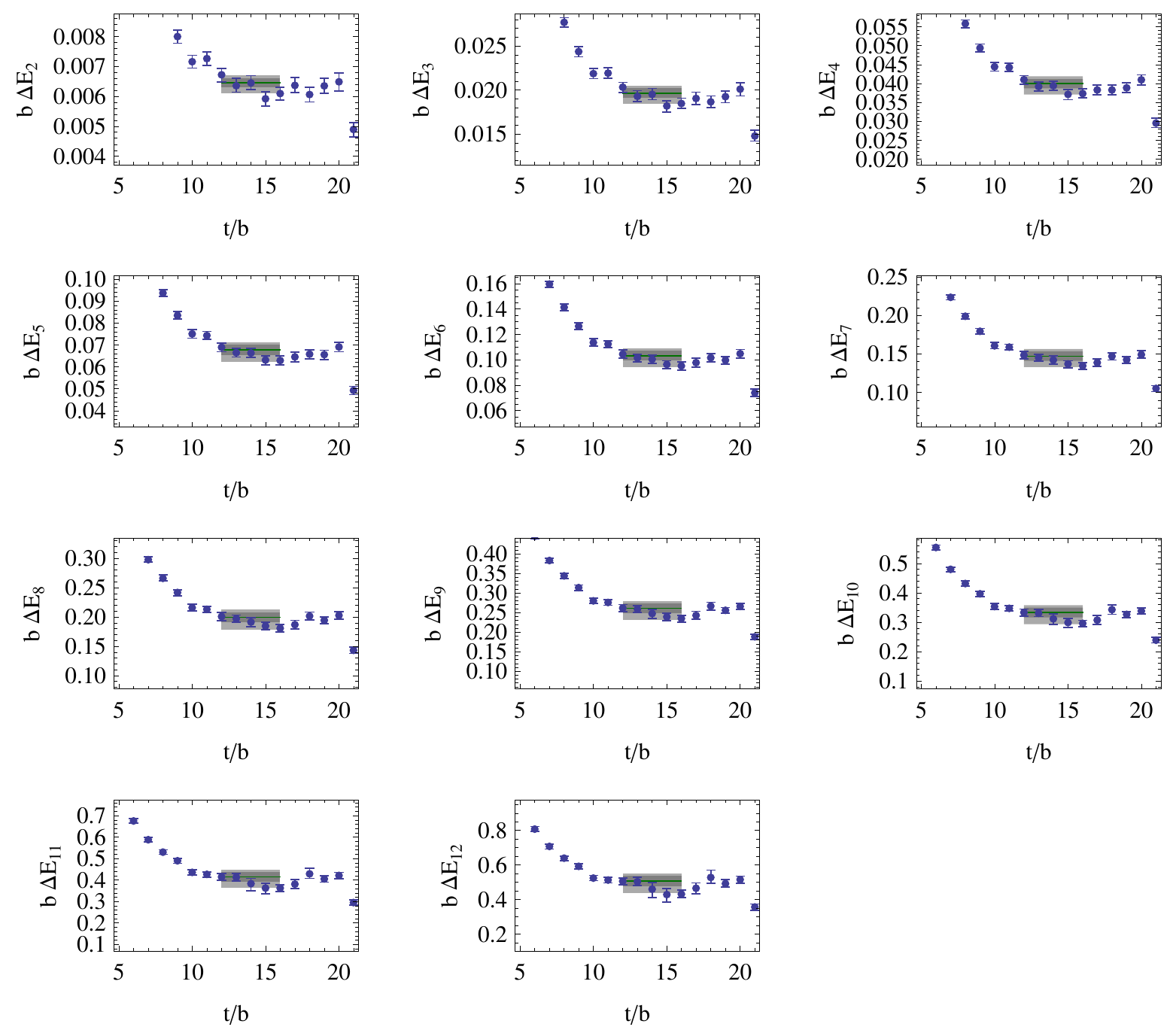}
  \caption{$n$-correlated fits to the $n=2,\ldots,12$ energy
    differences for the $m_\pi = 591$~MeV ensemble.}
  \label{fig:corr030}
\end{figure}

\begin{figure}[!t]
  \centering
  \includegraphics[width=0.90\columnwidth]{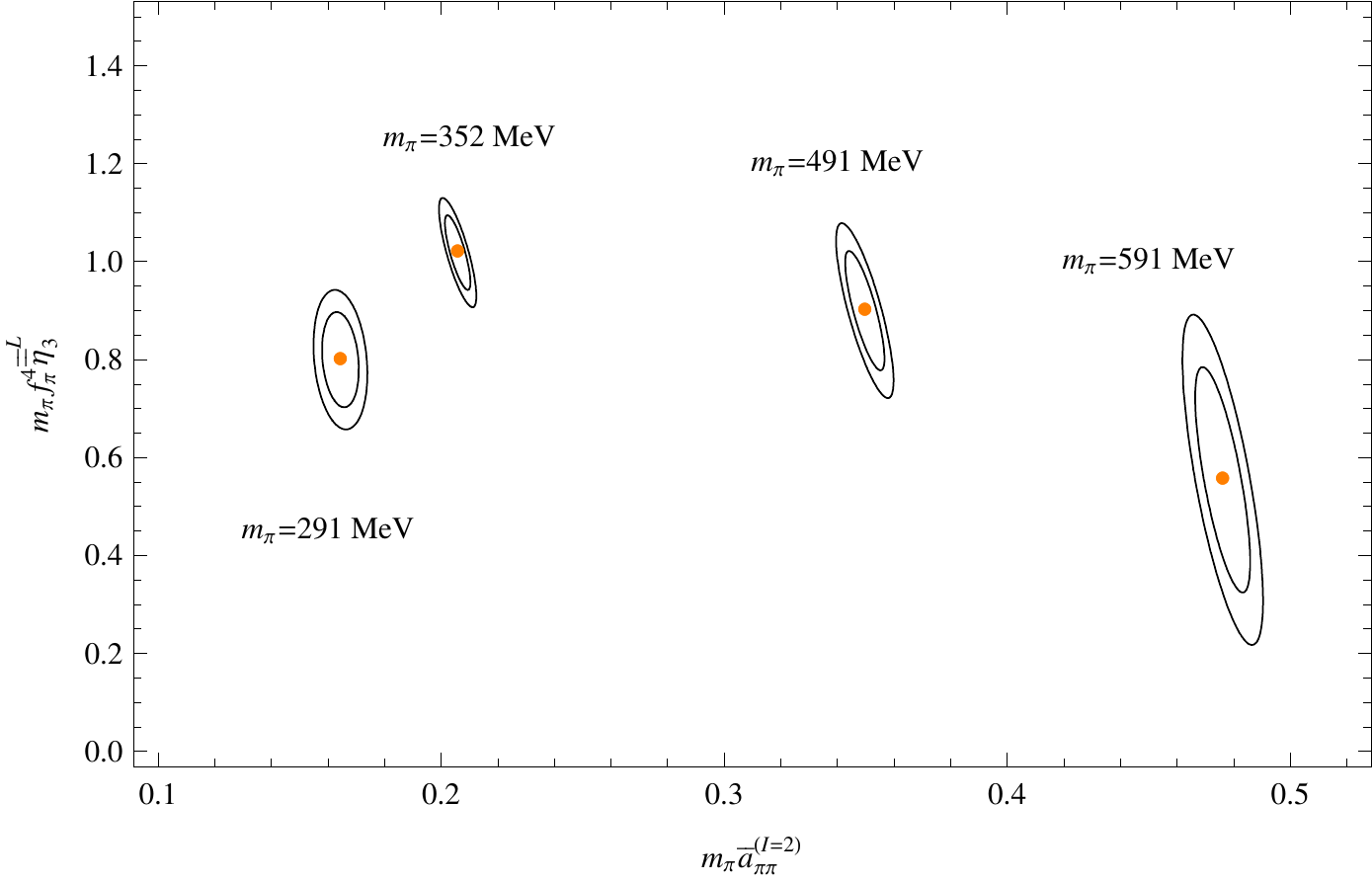}
  \caption{The $68\%$ and $90\%$ confidence regions for the two parameter fits performed to the
    $n=2,\ldots,12$ energy differences for each ensemble used in our
    work. Only statistical uncertainties are shown. }
  \label{fig:uncerelipse_summary}
\end{figure}

\noindent
The most complete use of the full set of energy differences that we
have computed is made by performing the coupled, ${\cal O}(L^{-7})$
analysis of the $n=2,\ldots,12$ effective energy differences to
extract $\apipi$ and \sheep, including the correlations in both $t$
and $n$ as discussed in the preceding section. The resultant fits of
such an analysis are shown in
figs.~\ref{fig:corr007}--\ref{fig:corr030} for the four ensembles. The
extracted fit parameters, $\apipi$ and \sheep, are central results of
this work.  They are given in Table \ref{table:abaretabarbarfits} and
their uncertainty ellipses are shown in
fig.~\ref{fig:uncerelipse_summary}.  For comparison with the simpler
analysis above, the shaded regions in fig.~\ref{fig:eta3bar_summary}
correspond to the values of \sheep\ extracted from this correlated
analysis and are seen to be consistent with the extractions made using
eq.~(\ref{eq:eta3barbarisolation}).  Similarly, the extracted $\apipi$
agrees with that obtained from the exact eigenvalue method for $n=2$
(and hence with all the N$^3$LO extractions above).
\begin{table}[ht]
\caption{The results of simultaneous fit of $\apipi$ and
  $\overline{\overline{\eta}}^L_3$
to all twelve correlation functions,
$n=2,3,.., 12$, on each  ensemble.
(Recall that the nuclear physics sign convention for $\apipi$ is being used.)
}
\label{table:abaretabarbarfits}
\resizebox{!}{1.2cm}{
\begin{tabular}{@{}|c | c | c | c | c |}
\hline
\  Quantity \ & 
\ \ \ $\qquad m_\pi\sim 291~{\rm MeV}\qquad$\ \ \ & 
\ \ \ $\qquad m_\pi\sim 352~{\rm MeV}\qquad$\ \ \ & 
\ \ \ $\qquad m_\pi\sim 491~{\rm MeV}\qquad$\ \ \ & 
\ \ \ $\qquad m_\pi\sim 591~{\rm MeV}\qquad$\ \  \  \\
\hline
$m_{\pi}/f_{\pi}$ & $1.990(11)(14)$ & 
        $2.3230(57)(30)$ & 
        $3.0585(49)(95)$ & 
        $3.4758(98)(60)$\\
\hline
$m_{\pi}\  \apipi$&
       $0.1644(40)\left({\tiny{
\begin{array}{c}+47\\-114\end{array}}}\right)$ &
       $0.2058(45)\left({\tiny{
\begin{array}{c}+46\\-82\end{array}}}\right)$ &
       $0.3497(69)\left({\tiny{
\begin{array}{c}+134\\-76\end{array}}}\right)$  &
       $0.4761(96)\left({\tiny{
\begin{array}{c}+126\\-198\end{array}}}\right)$\\
$m_{\pi}\ f_{\pi}^4\ \overline{\overline{\eta}}^L_3$\  &
       $0.80(09)\left({\tiny{
\begin{array}{c}+17\\-19\end{array}}}\right)$ &
       $1.02(08)\left({\tiny{
\begin{array}{c}+19\\-22\end{array}}}\right)$ &
       $0.90(12)\left({\tiny{
\begin{array}{c}+12\\-45\end{array}}}\right)$ &
       $0.55(23)\left({\tiny{
\begin{array}{c}+20\\-50\end{array}}}\right)$\\\hline
$\chi^2/{\rm dof}$\ ({\rm dof}) & 
        $1.3$ \ (65) & 
        $1.9$  \ (63) & 
        $1.8$  \ (63) &
        $1.3$ \ (53)  \\
\hline
\end{tabular}}
\end{table}

\begin{figure}[!t]
  \centering
  \includegraphics[width=0.90\columnwidth]{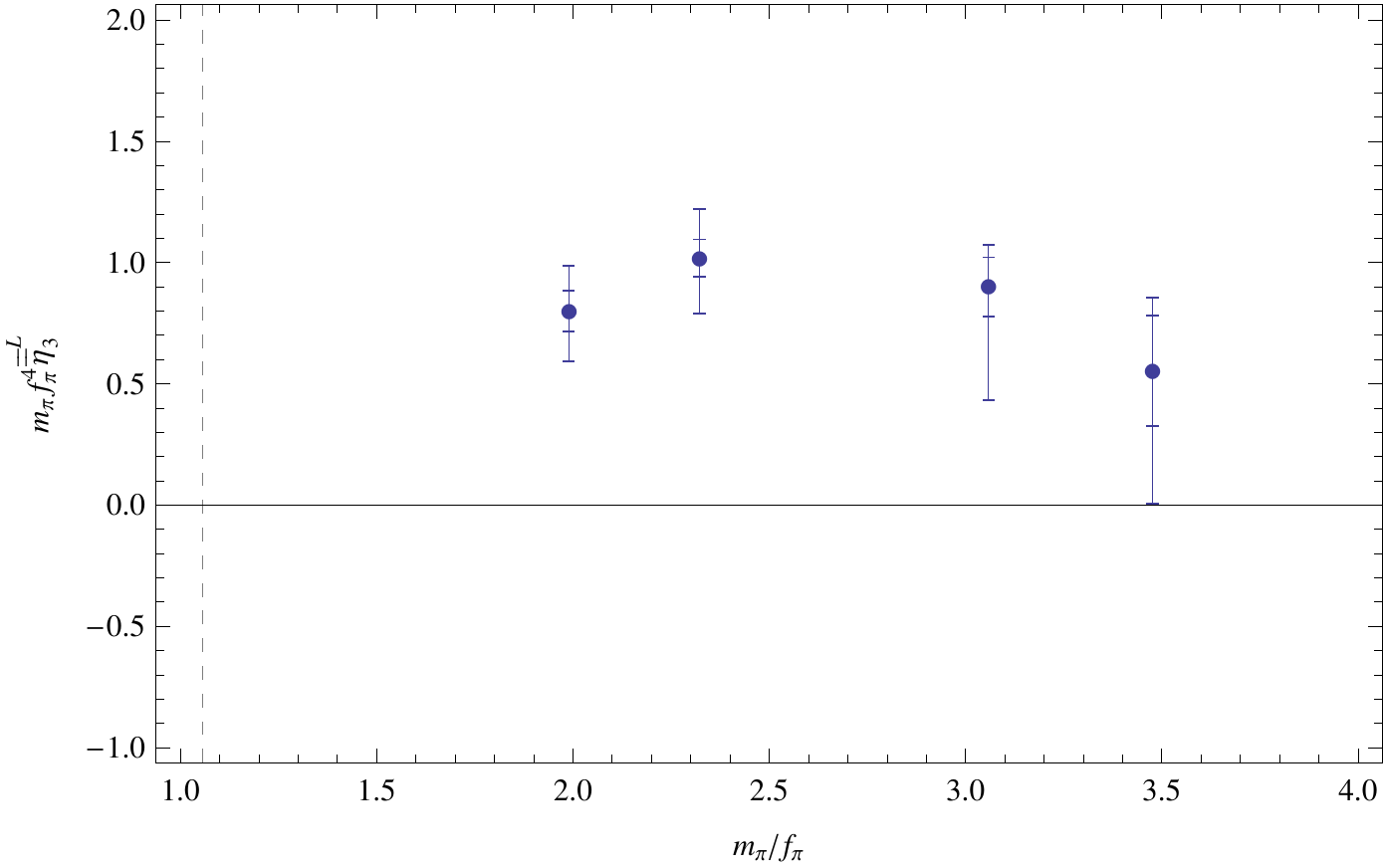}
  \caption{ The three-body interaction,  $m_{\pi} f_{\pi}^4\ 
    \overline{\overline{\eta}}^L_3$, defined in
    eq.~(\ref{eq:energyshift}), as a function of $m_{\pi}/f_{\pi}$.
    The Statistical uncertainties, and the statistical and systematic
    uncertainties combined in quadrature are shown.  The dashed
    vertical line indicates the physical point.}
  \label{fig:threebodysummary}
\end{figure}

In fig.~\ref{fig:threebodysummary}, \sheep\ is plotted versus
$m_{\pi}/f_{\pi}$, in units of the estimate based upon NDA,
$\overline{\overline{\eta}}^{L,NDA}_3 = 1/(m_\pi f_\pi^4)$, as
discussed in Ref.~\cite{Beane:2007qr,Detmold:2008gh}.  While the
three-body interaction is consistent with its NDA estimate for the
lightest three pion masses, it is found to be consistent with zero at
the heaviest mass, $m_\pi = 591~{\rm MeV}$.  It is desirable to reduce
the uncertainties in this calculation to see, if in fact, the
three-body interaction is decreasing with increasing pion mass.  If
this is found to be the case then a more detailed study in this high
pion mass region is warranted.

\begin{figure}[!t]
  \centering
  \includegraphics[width=0.9\columnwidth]{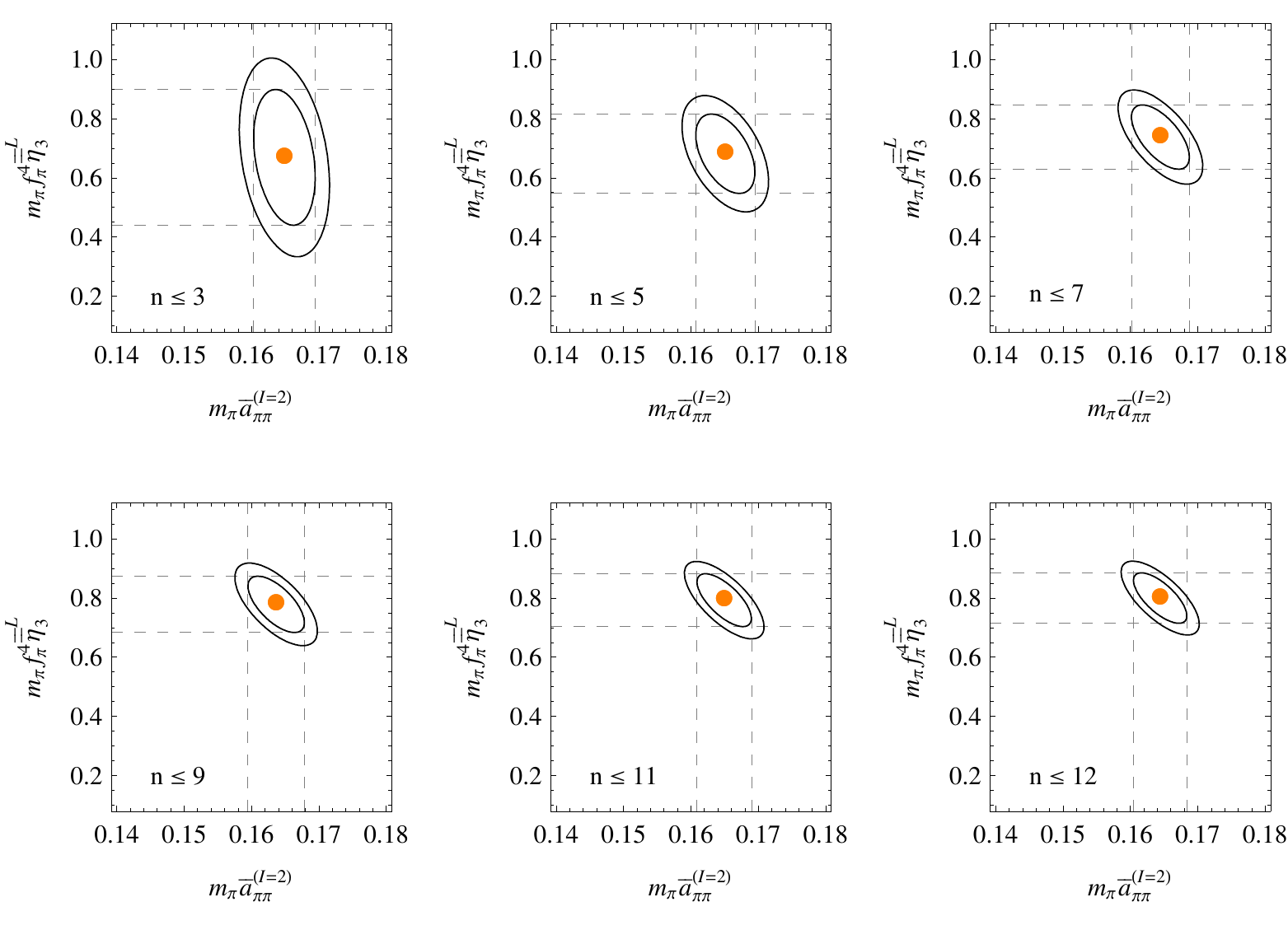}
  \caption{Dependence of the fit parameters, $\apipi$ and \sheep, on
    the number of energy differences included in the fit for the
    $m_\pi=291$~MeV ensemble. $n \leq n_{\rm max}$ implies that the
    energies up to $n_{\rm max}$ are included in the correlated fit.
    The inner and outer ellipses show the $68\%$ and $90\%$ confidence
    regions.  }
  \label{fig:fitvsnmax}
\end{figure}

It is interesting to study how the larger $n$ energy differences
influence the extraction of the two parameters. To do so, we have
performed a series of fits including only the energy differences up to
a given $n_{\rm max}$. The resulting confidence regions for the
parameters extracted from the $m_\pi = 291$~MeV ensemble are shown in
fig.~\ref{fig:fitvsnmax} for $n_{\rm max}=3,\ 5,\ 7,\ 9,\ 11,\ 12$.
Clearly the inclusion of higher $n$ data improves the determination of
\sheep\ in particular.

\section{The Three-Pion Interaction, $\overline{\overline{\eta}}^L_3$,
  $\overline{\eta}^L_3$, and $\eta_3 (\mu)$}
\label{sec:threebody}

\noindent
In the preceding Section the $\pi^+\pi^+\pi^+$-interaction, \sheep\ ,
has been extracted successfully from the lattice QCD calculations and
is non-zero at the lighter quark masses used in this study.  The
renormalization group invariant, but volume dependent three-body
interaction that arises at ${\cal O}(L^{-6})$, $\overline{\eta}^L_3$,
that receives perturbative corrections in the volume expansion to
become $\overline{\overline{\eta}}^L_3$ at ${\cal O}(L^{-7})$, is
somewhat more problematic to extract from these particular lattice
calculations due to the size of $\apipi$ in relation to the size of
the lattice.

There are two terms that must be calculated in order to determine
$\overline{\eta}^L_3$, one is additive and one is multiplicative, as
given in eq.~(\ref{eq:eta3barbar}).  Assuming that the effective range
for $\pi^+\pi^+$ scattering is not orders of magnitude larger than
$\apipi$, the additive term makes a very small contribution to
$\overline{\overline{\eta}}^L_3$.  To make this explicit, it is useful
to rewrite eq.~(\ref{eq:eta3barbar}) as
\begin{eqnarray}
&& m_{\pi}\ f_{\pi}^4\ \overline{\overline{\eta}}^L_3
 \ = \ 
\alpha_{\eta_3}\ m_{\pi}\ f_{\pi}^4\ \overline{\eta}^L_3
\ +\ 
\beta_{\eta_3}\ {r\over\abar}
\ \ \ ,
\label{eq:albedefn}
\end{eqnarray}
where
\begin{eqnarray}
&  & \alpha_{\eta_3} \ = \  1 \ -\  6  \left({\abar\over\pi L}\right)\ {\cal I} 
\ \ \ \ ,\ \ \ \ 
\beta_{\eta_3}\ =\ {72\pi \abar^5 \over  L}\  f_{\pi}^4\ {\cal I}
\ \ \ .
\label{eq:albedefnab}
\end{eqnarray}
The numerical values of $\beta_{\eta_3}$ are shown in
Table~\ref{table:albarvals}, and are all seen to be very small for
$r\sim| \abar |$.  The multiplicative factor, $\alpha_{\eta_3}$, is
also shown in Table~\ref{table:albarvals}, and its values make clear
that the perturbative expansion for the three-body interaction is
converging extremely slowly at the lightest pion mass, with a $\sim
75\%$ correction at ${\cal O}(L^{-7})$ to the ${\cal O}(L^{-6})$
result, and fails completely at both $m_\pi = 491~{\rm MeV}$ and
$591~{\rm MeV}$ for the lattice volumes used in this work.  Clearly
larger lattice volumes will be required in order to determine the
three-body interaction $\overline{\eta}^L_3$ with
precision.\footnote{We note that it is \sheep\ that was extracted in
  Ref.~\cite{Beane:2007es}.} This is in stark contrast to the two-body
scattering parameters which can be extracted to high precision from
these same lattice volumes and the perturbative expansion in
eq.~(\ref{eq:energyshift}) appears to be converging.
\begin{table}[ht]
\caption{Correction factors required to determine  $\overline{\eta}^L_3$ from
  $\overline{\overline{\eta}}^L_3$,
defined in eq.~(\protect\ref{eq:albedefnab}).
}
\label{table:albarvals}
\resizebox{!}{1.3cm}{
\begin{tabular}{@{}|c | c | c | c | c |}
\hline
\  Quantity \ & 
\ \ \ $\qquad m_\pi\sim 291~{\rm MeV}\qquad$\ \ \ & 
\ \ \ $\qquad m_\pi\sim 352~{\rm MeV}\qquad$\ \ \ & 
\ \ \ $\qquad m_\pi\sim 491~{\rm MeV}\qquad$\ \ \ & 
\ \ \ $\qquad m_\pi\sim 591~{\rm MeV}\qquad$\ \  \  \\
\hline
$m_{\pi}/f_{\pi}$ & $1.990(11)(14)$ & 
        $2.3230(57)(30)$ & 
        $3.0585(49)(95)$ & 
        $3.4758(98)(60)$\\
\hline
$\alpha_{\eta_3}$ & 
        $1.74(3)$ & 
        $1.78(2)$ & 
        $1.97(2)$  & 
        $2.08(3)$\\
$\beta_{\eta_3}$ & 
        $-0.0038(7)$ & 
        $-0.0056(7)$ & 
        $-0.020(2)$ &
        $-0.044(6)$\\
\hline
$\alpha_{\eta_3}$ $[ (3.5\ {\rm fm})^3 ]$& 
        $1.53(2)$ & 
        $1.56(1)$ & 
        $1.69(2)$  & 
        $1.77(2)$\\
$\beta_{\eta_3}$ $[ (3.5\ {\rm fm})^3 ]$ & 
        $-0.0027(5)$ & 
        $-0.0040(5)$ & 
        $-0.014(2)$ &
        $-0.031(4)$\\
\hline
\end{tabular}}
\end{table}
So while it is possible to determine the ``dressed'' three-pion
interaction, the uncertainty in the determination of the bare
three-pion interaction is large, not because of the uncertainty in the
lattice calculation, but due to the relatively large higher-order
terms in the volume expansion when evaluated at this present lattice
volume (a theoretical systematic uncertainty).  In
Table~\ref{table:albarvals}, the correction factors are also given for
a lattice of spatial volume (3.5 fm)$^3$ for comparison.  However,
even in these large volumes the correction factors at one higher order
in the large volume expansion of $\sim 50\%$ and clearly even larger
volumes, $L \gsim 3.5~{\rm fm}$, are required in order to have the
volume expansion of the three-pion interaction under perturbative
control.

The quantity $\eta_3(\mu)$ is a renormalization scheme dependent
quantity that is independent of the volume, and as such is the
quantity that most directly enters into the calculation of other
many-body processes.  It is easily extracted from
$\overline{\eta}_3^L$ via eq.~(\ref{eq:etathreebar}).  However, given
the present theoretical systematic uncertainties in
$\overline{\eta}_3^L$, we do not attempt a determination of
$\eta_3(\mu)$.

\section{The Equation of State and the Isospin Chemical Potential}
\label{sec:chemical}

\begin{figure}[!t]
  \centering
  \includegraphics[width=0.90\columnwidth]{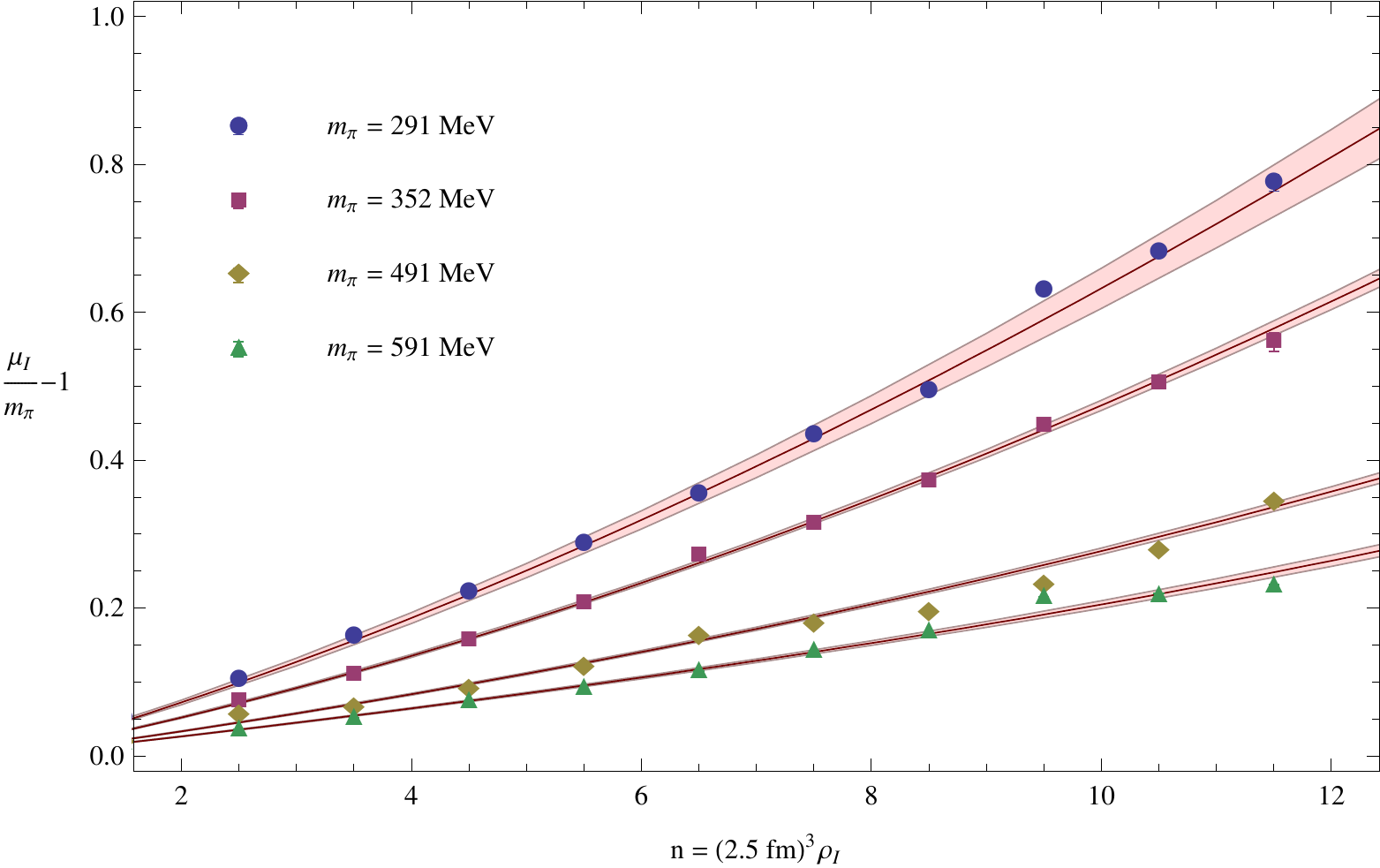}
  \caption{The isospin chemical potential as a function of the number of pions
    in a fixed volume (equivalent to the isospin density $\rho_I$) .
    The points and their associated uncertainties are the results of
    the lattice calculation where a finite-difference has been used to
    construct the derivative with respect to the number of $\pi^+$'s.
    The solid curves and bands result from the analytic expression for
    the energy of the ground state in the large volume expansion,
    eq.~(\ref{eq:energyshift}), using the fit values for $\apipi$ and
    \sheep\ and their correlated uncertainties.  }
\label{fig:isospin}
\end{figure}

\noindent
The energy of the $n$-$\pi^+$ system as a function of volume and of
the number of $\pi^+$'s is given explicitly in
eq.~(\ref{eq:energyshift}) in the large-volume expansion.  From the
equation of state, the isospin chemical potential is defined as
\begin{eqnarray}
  \label{eq:isospindensity}
  \mu_I &=& \left.\frac{d E_n}{d n}\right|_{V={\rm const}} 
  \ \ \ ,
\end{eqnarray}
which can be constructed analytical from eq.~(\ref{eq:energyshift}) or
numerically from the results of the lattice calculation by using a
simple finite difference approximation.  The resulting ratio of this
isospin-chemical potential to the pion mass for each of our ensembles
is shown in fig.~\ref{fig:isospin} as a function of $n$, and for
convenience, the isospin density, $\rho_I$ (which in this case is the
number density).  We note that for the $12$-$\pi^+$ system, the number
density is $\rho_I^{(12)}=12/L^3 = 0.77~{\rm fm}^{-3}$.  In
fig.~\ref{fig:isospin}, the solid curves corresponds to the prediction
at ${\cal O}(L^{-7})$ from eq.~(\ref{eq:energyshift}), and this
prediction differs insignificantly from that at ${\cal O}(L^{-6})$.
There have been recent works that perform lattice QCD calculations at
finite isospin chemical potential, e.g.
Ref.~\cite{Kogut:2002se,deForcrand:2007uz}, where the starting point
is a chemical potential term for the quarks added to the QCD action.
Compelling results for the formation of the charged-pion condensate at
$\mu_I\sim m_\pi$ have been produced.

\begin{figure}[!t]
  \centering
  \includegraphics[width=0.90\columnwidth]{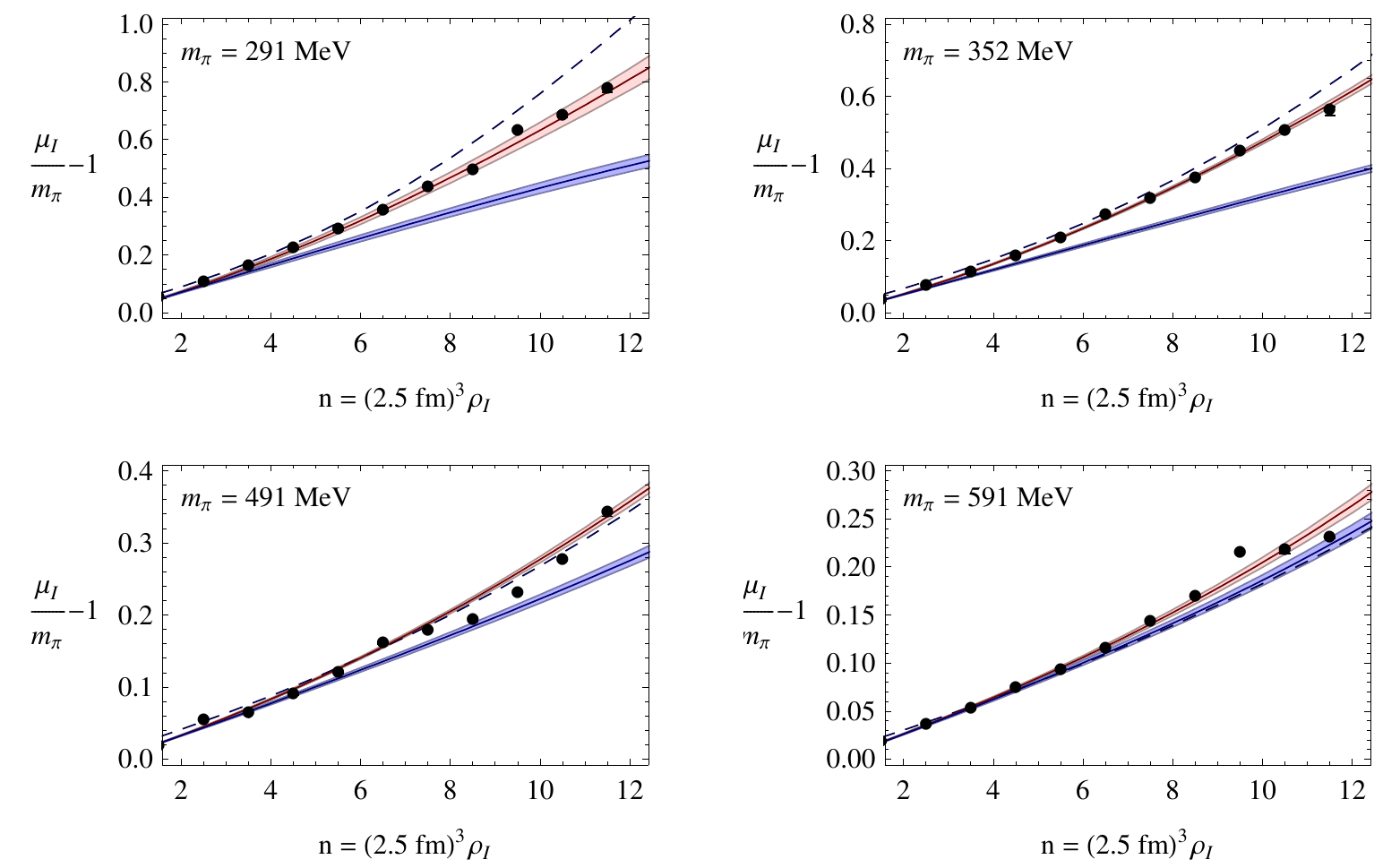}
  \caption{
    The isospin chemical potential as a function of the number of
    pions (equivalent to the isospin density $\rho_I$) at a fixed
    volume with and without the contribution from the
    $\pi^+\pi^+\pi^+$-interaction, \sheep.  The solid red (lighter)
    curves and bands result from the analytic expression for the
    energy of the ground state in the large volume expansion,
    eq.~(\ref{eq:energyshift}), using the fit values for $\apipi$ and
    \sheep and their correlated uncertainties. The solid blue (darker)
    curves and bands are similarly the results for the fitted value of
    $\apipi$ and \sheep=0. The dashed curve corresponds to the leading
    order prediction of $\chi$PT.  }
\label{fig:isospinNothreebody}
\end{figure}

Pionic systems at finite isospin chemical potential (and temperature)
have been explored theoretically in
$\chi$PT~\cite{Kogut:1999iv,Kogut:2000ek,Son:2000xc,Son:2000by},
primarily as a step toward understanding finite density nuclear
systems.  The inclusion of the isospin chemical potential can be
accomplished straightforwardly by replacing the time-component of the
covariant derivative acting on the exponential field of $\pi$'s,
$D_0\Sigma $, with $\nabla_0\Sigma = D_0\Sigma - i \mu_I {1\over
  2}\left[ \tau_3\ ,\ \Sigma\ \right]$.  For $\mu_I < m_\pi$ the
vacuum state is the same as it is for $\mu_I=0$, $\Sigma=1$, but for
$\mu_I > m_\pi$ the vacuum alignment changes and becomes,
\begin{eqnarray}
\overline{\Sigma} & = & \cos\alpha\ +\ i\left( \tau_1\cos\phi + \tau_2\sin\phi
  \right)\sin\alpha
\ \ \ ,
\end{eqnarray}
using the notation of Refs.~\cite{Son:2000xc,Son:2000by}.  Minimizing
the potential energy at LO in $\chi$PT gives $\cos\alpha =
m_\pi^2/\mu_I^2$ and a relation between the isospin density, $\rho_I$,
and chemical potential~\footnote{The numerical factors that appear in
  eq.~(\ref{eq:LOchiPTmuI}) differ by factors of two from
  Refs.~\cite{Kogut:1999iv,Kogut:2000ek,Son:2000xc,Son:2000by} due to
  the definition of $f_\pi$.  At the physical pion mass we use
  $f_\pi\sim 132~{\rm MeV}$.},
\begin{eqnarray}
\rho_I & = & {1\over 2}\ f_\pi^2\ \mu_I\ \left(\ 1 - {m_\pi^4\over\mu_I^4}\ \right)
\ =\ 2 f_\pi^2\ \left(\mu_I-m_\pi\right)\ -\ 3{f_\pi^2\over m_\pi} \
\left(\mu_I-m_\pi\right)^2\ +\ ...
\ \ \ .
\label{eq:LOchiPTmuI}
\end{eqnarray}
By construction, our lattice QCD calculation is in the regime of
$\mu_I > m_\pi$, and we expect that these LO $\chi$PT results should
come perturbatively close to describing the properties of the
$n$-$\pi^+$ systems.  Implicit in this LO $\chi$PT analysis are not
only the $\pi^+\pi^+$-interactions, but also multi-pion interactions,
including the $\pi^+\pi^+\pi^+$-interaction.  In fact, the NDA
estimate of the $\pi^+\pi^+\pi^+$-interaction arises from the LO terms
in $\chi$PT.  In fig.~\ref{fig:isospinNothreebody}, we show the
isospin chemical potential versus the number of $\pi^+$'s in the
lattice volume, which can be directly translated into isospin density,
$\rho_I=n/L^3$. To determine the importance of the
$\pi^+\pi^+\pi^+$-interaction, we show the $\mu_I$ calculated directly
from the lattice calculation. We also show the curve resulting from
eq.~(\ref{eq:energyshift}) with the values of $\apipi$ and \sheep\ 
extracted from the lattice calculation and their associated correlated
uncertainties (red curves and shaded regions), and we show the curve
resulting from setting \sheep$=0$ in eq.~(\ref{eq:energyshift}) (blue
curve and shaded region).  It is clear from
fig.~\ref{fig:isospinNothreebody} that the
$\pi^+\pi^+\pi^+$-interaction plays an important role in the relation
between the isospin chemical potential and the isospin density.
Further, in fig.~\ref{fig:isospinNothreebody}, the dashed curve is the
result of LO $\chi$PT, as given in eq.~(\ref{eq:LOchiPTmuI}), which is
seen to describe the result of the lattice calculation well at all the
pion masses we have explored.  There does appear to be a slight
$m_\pi$-dependent systematic difference, but the magnitude of this
effect is consistent with terms higher order in $\chi$PT.

Another quantity of interest that one wishes to determine for such
systems is the pressure as a function of the $\rho_I$,
\begin{eqnarray}
  \label{eq:pressure}
P & = &  \left. {1\over 3 L^2} {d E_n\over d L}\right|_{n={\rm const}}
  \ \ \ .
\end{eqnarray}
Unfortunately, we cannot recover the pressure directly from our
lattice calculations using eq.~(\ref{eq:energyshift}) because both
$\apipi$ and \sheep\ depend implicitly upon the volume. Further,
without lattice calculations of different volumes, the pressure cannot
be determined directly from the lattice calculations either.
Therefore, with the present lattice calculations we have performed,
while the isospin chemical potential can be determined as a function
of density, the pressure cannot.

An important issue to consider is the impact of the finite lattice
volume upon the relation between the isospin chemical potential and
the isospin density.  The periodic boundary conditions imposed in the
spatial directions leave the zero-mode untouched, but discretize the
non-zero modes.  As such, the tree-level matrix element of the
interaction Hamiltonian between initial and final states in which each
pion carries zero three-momentum is unaffected by the finite volume.
However, the boundary conditions will modify contributions at the
one-loop level and beyond.  Therefore, in the limit where the
contributions from loop-level diagrams are small, as is the case for a
small scattering length, the modifications of the relation between
isospin chemical potential and isospin density due to the boundary
conditions is expected to be perturbatively small\footnote{ In
  infinite volume, the leading contributions to the ground state
  energy of an imperfect dilute Bose gas are calculated easily by
  performing a Bogoliubov transformation on the Hamiltonian, leaving
  the leading order mean-field contribution and a term resulting from
  a summation over the non-zero momentum states.  The leading terms in
  the density expansion of the energy per particle on the ground-state
  are
\begin{eqnarray}
E/N & = & {2\pi a\rho\over M}\left[\ 1\ +\ {128\over 15}\ \sqrt{\rho
    a^3\over\pi}\ +\ ...\right]
\ \ \ .
\label{eq:densityexpansion}
\end{eqnarray}
When placed in a finite cubic volume with periodic boundary
conditions, the integral over non-zero momentum states that gives rise
to the term that is non-analytic in the number-density, $\rho$, is
replaced by a summation over the allowed momentum states in the
volume.  When the unperturbed level spacing in the volume is large
compared with the energy-shift at leading order in the density
expansion, the contents of the square brackets in
eq.~(\ref{eq:densityexpansion}) becomes
\begin{eqnarray}
\left[\ 1\ +\ {128\over 15}\ \sqrt{\rho
    a^3\over\pi}\ +\ ...\right]
&\rightarrow &
\left[\ 1\ -\ \left({a\over \pi L}\right) \ {\cal I}\ +\ ... \right]
\ \ \ ,
\label{eq:densitygoesto}
\end{eqnarray}
recovering the leading terms in the finite-volume expansion in
eq.~(\ref{eq:energyshift}) as the number of particles becomes large.
See also Ref.~\cite{Tan:2007bg}.  }.  In systems with large scattering
lengths, e.g. nucleons, the finite-volume modifications may be
substantial, and this requires further study.

\section{Conclusions}
\label{sec:disc}
\noindent
One of the major challenges facing nuclear physics is the solution of
the hadronic many-body problem, including the effects of the
multi-hadron interactions induced at the scale of chiral symmetry
breaking.  While the two-nucleon interaction is overwhelmingly the
dominant interaction in multi-nucleon systems, a three-nucleon
interaction contributes significantly to the structure and
interactions of nuclei.  Calculating the properties and interactions
of nuclei is a major goal of lattice QCD and as a small step in this
direction we have performed the first lattice QCD calculations of the
properties of systems comprised of multiple-$\pi^+$'s.  The present
paper is a detailed follow-up to the letter we recently
published~\cite{Beane:2007es}, and extends that work from studies of
systems comprised of up to five pions, to systems containing up to
twelve pions. As a technical detail, this required calculations of
very high (64 decimal digit) precision.

We have convincingly determined a non-zero value for the dressed
three-body interaction, $\overline{\overline{\eta}}_3^L$, defined in
eq.~(\ref{eq:energyshift}), at the lightest three pion masses, $m_\pi
= 291, 352$ and $491~{\rm MeV}$, and find a value consistent with zero
at the heaviest mass, $m_\pi = 591~{\rm MeV}$. These central results
are summarized in Table~\ref{table:abaretabarbarfits} and
fig.\ref{fig:threebodysummary} above.  Given the lattice volumes in
which the calculations have been performed, and the value of the
$\pi^+ \pi^+$ scattering length, the connection between \sheep\ and
the underlying three-body interaction, $\eta_3(\mu)$, (see
eqs.~(\ref{eq:eta3barbar}) and (\ref{eq:etathreebar})) is slowly
converging, making a meaningful extraction of $\eta_3(\mu)$ currently
impractical.  Clearly, in addition to higher order calculations
(${\cal O}(L^{-8})$ and beyond) of the energy of multi-pion systems in
a finite volume, lattice calculations in larger volumes, and of
excited states in these volumes, are required to disentangle the bare
from the dressed interactions.

An important outcome of our calculations is a determination of the
relation between the isospin chemical potential and the isospin
density.  For $m_\pi\lsim 400~{\rm MeV}$ the $\pi^+\pi^+\pi^+$
interaction is seen to make a sizable contribution to this relation.
Such contributions are implicit in the LO $\chi$PT theoretical result,
but our calculation has provided the first explicit QCD calculation of
the importance of multi-pion interactions.

This work and our previous Letter represent the first study of
multi-hadrons systems directly from QCD. Whilst encouraging, we
conclude by reiterating that a significant amount of theoretical work
remains to be performed in order to explore meaningfully more
complicated systems such as those involving large numbers of baryons.

\section{Acknowledgments}

\noindent We thank B.~Bringoltz M.~Endres, D.~B.~Kaplan and D.~T.~Son
for discussions.  We thank R.~Edwards and B.~Joo for help with the
QDP++/Chroma programming environment~\cite{Edwards:2004sx} with which
the calculations discussed here were performed.  The computations for
this work were performed at Jefferson Lab, Fermilab, Lawrence
Livermore National Laboratory, National Center for Supercomputing
Applications, Centro Nacional de Supercomputaci\'on (Barcelona, Spain)
and the Institute for Nuclear Theory.  We are indebted to the MILC and
the LHP collaborations for use of their configurations and
propagators, respectively.  The work of MJS and WD was supported in
part by the U.S.~Dept.~of Energy under Grant No.~DE-FG03-97ER4014. The
work of KO was supported in part by the U.S.~Dept.~of Energy contract
No.~DE-AC05-06OR23177 (JSA) and by the Jeffress Memorial Trust, grant
J-813, DOE OJI grant DE-FG02-07ER41527 and DOE grant
DE-FG02-04ER41302.  The work of SRB and AT was supported in part by
the National Science Foundation CAREER grant No.  PHY-0645570. Part of
this work was performed under the auspices of the US DOE by the
University of California, Lawrence Livermore National Laboratory under
Contract No. W-7405-Eng-48.  The work of AP was partly supported by
the EU contract FLAVIAnet MRTN-CT-2006-035482, by the contract
FIS2005-03142 from MEC (Spain) and FEDER and by the Generalitat de
Catalunya contract 2005SGR-00343.

\appendix

\section{Contractions for $I_z=n\leq13$ Mesons }
\label{sec:contr-part}

\noindent
The contractions required for the $n=1$,\ldots, 13 correlation
functions are given below. The notation
\begin{eqnarray*}
  T_j &=& {\rm tr_{C,S}}\left[\Pi^j\right]
\end{eqnarray*}
is used for brevity with the trace being over color and spinor
indices.  The index, $j$, is the matrix power to which $\Pi$ is
raised.  {\footnotesize
  \begin{eqnarray}
    \label{eq:20}
    C_1(t) &=& T_1\,,
  \end{eqnarray}
  \begin{eqnarray}
    \label{eq:19}
        C_2(t) &=& T_1^2 - T_2\,,
  \end{eqnarray}
  \begin{eqnarray}
    \label{eq:18}
        C_3(t) &=& T_1^3-3 T_2 T_1+2 T_3\,,
  \end{eqnarray}
  \begin{eqnarray}
    \label{eq:17}
        C_4(t) &=& T_1^4-6 T_2 T_1^2+8 T_3 T_1+3 T_2^2-6 T_4\,,
  \end{eqnarray}
  \begin{eqnarray}
    \label{eq:16}
        C_5(t) &=& T_1^5-10 T_2 T_1^3+20 T_3 T_1^2+15 T_2^2 T_1
        -30 T_4 T_1-20 T_2 T_3+24 T_5\,,
  \end{eqnarray}
  \begin{eqnarray}
    \label{eq:15}
        C_6(t) &=& T_1^6-15 T_2 T_1^4+40 T_3 T_1^3+45 T_2^2 T_1^2
        -90 T_4 T_1^2-120 T_2 T_3 T_1
\nonumber\\&&
+144 T_5 T_1-15 T_2^3+40 T_3^2
        +90 T_2 T_4-120 T_6\,,
  \end{eqnarray}
  \begin{eqnarray}
    \label{eq:14}
        C_7(t) &=& T_1^7-21 T_2 T_1^5+70 T_3 T_1^4+105 T_2^2 T_1^3
        -210 T_4 T_1^3-420 T_2 T_3 T_1^2+504 T_5 T_1^2-105 T_2^3 T_1
\nonumber\\&&
+280 T_3^2 T_1+630 T_2 T_4 T_1-840 T_6 T_1+210 T_2^2 T_3
        -420 T_3 T_4-504 T_2 T_5+720 T_7\,,
  \end{eqnarray}
  \begin{eqnarray}
    \label{eq:13}
        C_8(t) &=& T_1^8-28 T_2 T_1^6+112 T_3 T_1^5+210 T_2^2 T_1^4
-420 T_4 T_1^4-1120 T_2 T_3 T_1^3+1344 T_5 T_1^3-420 T_2^3 T_1^2
\nonumber\\&&
+1120 T_3^2 T_1^2+2520 T_2 T_4 T_1^2-3360 T_6 T_1^2+1680 T_2^2 T_3 T_1
-3360 T_3 T_4 T_1-4032 T_2 T_5 T_1
+5760 T_7 T_1
\nonumber\\&&
+105 T_2^4-1120
   T_2 T_3^2+1260 T_4^2-1260 T_2^2 T_4+2688 T_3 T_5+3360 T_2 T_6-5040 T_8\,,
  \end{eqnarray}
  \begin{eqnarray}
    \label{eq:12}
    C_9(t) &=& T_1^9-36 T_2 T_1^7+168 T_3 T_1^6+378 T_2^2 T_1^5-756 T_4 T_1^5-2520 T_2 T_3 T_1^4+3024 T_5 T_1^4-1260 T_2^3 T_1^3
\nonumber\\&&
+3360 T_3^2
   T_1^3+7560 T_2 T_4 T_1^3-10080 T_6 T_1^3+7560 T_2^2 T_3 T_1^2-15120
   T_3 T_4 T_1^2-18144 T_2 T_5 T_1^2
\nonumber\\&&
+25920 T_7 T_1^2+945
   T_2^4 T_1
-10080 T_2 T_3^2 T_1+11340 T_4^2 T_1-11340 T_2^2 T_4 T_1+24192 T_3 T_5
T_1
\nonumber\\&&
+30240 T_2 T_6 T_1-45360 T_8 T_1+2240
   T_3^3-2520 T_2^3 T_3+15120 T_2 T_3 T_4+9072 T_2^2 T_5-18144 T_4 T_5
\nonumber\\&&
-20160 T_3 T_6-25920 T_2 T_7+40320 T_9\,,
  \end{eqnarray}
  \begin{eqnarray}
    \label{eq:11}
    C_{10}(t) &=& T_1^{10}-45 T_2 T_1^8+240 T_3 T_1^7+630 T_2^2
    T_1^6-1260 T_4 T_1^6-5040 T_2 T_3 T_1^5+6048 T_5 T_1^5-3150 T_2^3
    T_1^4 
\nonumber\\&&
+8400
   T_3^2 T_1^4+18900 T_2 T_4 T_1^4-25200 T_6 T_1^4+25200 T_2^2 T_3
   T_1^3-50400 T_3 T_4 T_1^3-60480 T_2 T_5 T_1^3 
\nonumber\\&&
+86400 T_7
   T_1^3+4725 T_2^4 T_1^2-50400 T_2 T_3^2 T_1^2+56700 T_4^2
   T_1^2-56700 T_2^2 T_4 T_1^2+120960 T_3 T_5 T_1^2 
\nonumber\\&&
+151200 T_2 T_6
   T_1^2-226800 T_8 T_1^2+22400 T_3^3 T_1-25200 T_2^3 T_3 T_1+151200
   T_2 T_3 T_4 T_1+90720 T_2^2 T_5 T_1 
\nonumber\\&&
-181440 T_4 T_5
   T_1-201600 T_3 T_6 T_1-259200 T_2 T_7 T_1+403200 T_9 T_1-945
   T_2^5+25200 T_2^2 T_3^2 
\nonumber\\&&
-56700 T_2 T_4^2+72576 T_5^2+18900
   T_2^3 T_4-50400 T_3^2 T_4-120960 T_2 T_3 T_5-75600 T_2^2 T_6
\nonumber\\&&
+151200 T_4 T_6+172800 T_3 T_7+226800 T_2 T_8-362880 T_{10}\,,
  \end{eqnarray}
  \begin{eqnarray}
    \label{eq:9}
  C_{11}(t) &=&    T_1^{11}-55 T_2 T_1^9+330 T_3 T_1^8+990 T_2^2 T_1^7-1980 T_4 T_1^7-9240 T_2 T_3 T_1^6+11088 T_5 T_1^6-6930 T_2^3
   T_1^5
\nonumber\\&&
+18480 T_3^2 T_1^5+41580 T_2 T_4 T_1^5-55440 T_6 T_1^5+69300 T_2^2 T_3 T_1^4-138600 T_3 T_4 T_1^4-166320 T_2 T_5
   T_1^4
\nonumber\\&&
+237600 T_7 T_1^4+17325 T_2^4 T_1^3-184800 T_2 T_3^2 T_1^3+207900 T_4^2 T_1^3-207900 T_2^2 T_4 T_1^3+443520 T_3 T_5
   T_1^3
\nonumber\\&&
+554400 T_2 T_6 T_1^3-831600 T_8 T_1^3+123200 T_3^3 T_1^2-138600 T_2^3 T_3 T_1^2+831600 T_2 T_3 T_4 T_1^2
\nonumber\\&&
+498960 T_2^2
   T_5 T_1^2-997920 T_4 T_5 T_1^2-1108800 T_3 T_6 T_1^2-1425600 T_2 T_7 T_1^2+2217600 T_9 T_1^2
\nonumber\\&&
-10395 T_2^5 T_1+277200 T_2^2
   T_3^2 T_1-623700 T_2 T_4^2 T_1+798336 T_5^2 T_1+207900 T_2^3 T_4 T_1-554400 T_3^2 T_4 T_1
\nonumber\\&&
-1330560 T_2 T_3 T_5 T_1-831600
   T_2^2 T_6 T_1+1663200 T_4 T_6 T_1+1900800 T_3 T_7 T_1+2494800 T_2 T_8 T_1
\nonumber\\&&
-3991680 T_{10} T_1-123200 T_2 T_3^3+415800 T_3
   T_4^2+34650 T_2^4 T_3-415800 T_2^2 T_3 T_4-166320 T_2^3 T_5
\nonumber\\&&
+443520 T_3^2 T_5+997920 T_2 T_4 T_5+1108800 T_2 T_3 T_6-1330560
   T_5 T_6+712800 T_2^2 T_7-1425600 T_4 T_7
\nonumber\\&&
-1663200 T_3 T_8-2217600 T_2 T_9+3628800 T_{11}\,,
  \end{eqnarray}
  \begin{eqnarray}
    \label{eq:8}
  C_{12}(t) &=&    T_1^{12}-66 T_2 T_1^{10}+440 T_3 T_1^9+1485 T_2^2 T_1^8-2970 T_4 T_1^8-15840 T_2 T_3 T_1^7+19008 T_5 T_1^7-13860 T_2^3
   T_1^6
\nonumber\\&&
+36960 T_3^2 T_1^6+83160 T_2 T_4 T_1^6-110880 T_6 T_1^6+166320 T_2^2 T_3 T_1^5-332640 T_3 T_4 T_1^5-399168 T_2 T_5
   T_1^5
\nonumber\\&&
+570240 T_7 T_1^5+51975 T_2^4 T_1^4-554400 T_2 T_3^2 T_1^4+623700 T_4^2 T_1^4-623700 T_2^2 T_4 T_1^4+1330560 T_3 T_5
   T_1^4
\nonumber\\&&
+1663200 T_2 T_6 T_1^4-2494800 T_8 T_1^4+492800 T_3^3 T_1^3-554400 T_2^3 T_3 T_1^3+3326400 T_2 T_3 T_4 T_1^3
\nonumber\\&&
+1995840
   T_2^2 T_5 T_1^3-3991680 T_4 T_5 T_1^3-4435200 T_3 T_6 T_1^3-5702400 T_2 T_7 T_1^3+8870400 T_9 T_1^3
\nonumber\\&&
-62370 T_2^5
   T_1^2+1663200 T_2^2 T_3^2 T_1^2-3742200 T_2 T_4^2 T_1^2+4790016 T_5^2 T_1^2+1247400 T_2^3 T_4 T_1^2
\nonumber\\&&
-3326400 T_3^2 T_4
   T_1^2-7983360 T_2 T_3 T_5 T_1^2-4989600 T_2^2 T_6 T_1^2+9979200 T_4 T_6 T_1^2+11404800 T_3 T_7 T_1^2
\nonumber\\&&
+14968800 T_2 T_8
   T_1^2-23950080 T_{10} T_1^2-1478400 T_2 T_3^3 T_1+4989600 T_3 T_4^2 T_1+415800 T_2^4 T_3 T_1
\nonumber\\&&
-4989600 T_2^2 T_3 T_4
   T_1-1995840 T_2^3 T_5 T_1+5322240 T_3^2 T_5 T_1+11975040 T_2 T_4 T_5 T_1+13305600 T_2 T_3 T_6 T_1
\nonumber\\&&
-15966720 T_5 T_6
   T_1+8553600 T_2^2 T_7 T_1-17107200 T_4 T_7 T_1-19958400 T_3 T_8 T_1-26611200 T_2 T_9 T_1
\nonumber\\&&
+43545600 T_{11} T_1+10395
   T_2^6+246400 T_3^4-1247400 T_4^3-554400 T_2^3 T_3^2+1871100 T_2^2 T_4^2
\nonumber\\&&
-4790016 T_2 T_5^2+6652800 T_6^2-311850 T_2^4
   T_4+3326400 T_2 T_3^2 T_4+3991680 T_2^2 T_3 T_5-7983360 T_3 T_4 T_5
\nonumber\\&&
+1663200 T_2^3 T_6-4435200 T_3^2 T_6-9979200 T_2 T_4
   T_6-11404800 T_2 T_3 T_7
+13685760 T_5 T_7\nonumber\\&&
-7484400 T_2^2 T_8+14968800 T_4 T_8+17740800 T_3 T_9
+23950080 T_2 T_{10}-39916800   T_{12}\,,
  \end{eqnarray}

\begin{eqnarray}
  \label{eq:7}
  C_{13}(t) &=&
T_1^{13}-78 T_2 T_1^{11}+572 T_3 T_1^{10}+2145 T_2^2 T_1^9-4290 T_4
T_1^9-25740 T_2 T_3 T_1^8+30888 T_5 T_1^8
\nonumber\\&&
-25740
   T_2^3 T_1^7+68640 T_3^2 T_1^7+154440 T_2 T_4 T_1^7-205920 T_6 T_1^7+360360 T_2^2 T_3 T_1^6
\nonumber\\&&
-720720 T_3 T_4 T_1^6-864864 T_2
   T_5 T_1^6+1235520 T_7 T_1^6+135135 T_2^4 T_1^5-1441440 T_2 T_3^2 T_1^5
\nonumber\\&&
+1621620 T_4^2 T_1^5-1621620 T_2^2 T_4 T_1^5+3459456
   T_3 T_5 T_1^5+4324320 T_2 T_6 T_1^5-6486480 T_8 T_1^5
\nonumber\\&&
+1601600 T_3^3 T_1^4
-1801800 T_2^3 T_3 T_1^4+10810800 T_2 T_3 T_4
   T_1^4+6486480 T_2^2 T_5 T_1^4-12972960 T_4 T_5 T_1^4
\nonumber\\&&
-14414400 T_3 T_6 T_1^4
-18532800 T_2 T_7 T_1^4+28828800 T_9
   T_1^4-270270 T_2^5 T_1^3
+7207200 T_2^2 T_3^2 T_1^3
\nonumber\\&&
-16216200 T_2 T_4^2 T_1^3+20756736 T_5^2
T_1^3+5405400 T_2^3 T_4 
   T_1^3
-14414400 T_3^2 T_4 T_1^3
-34594560 T_2 T_3 T_5 T_1^3
\nonumber\\&&
-21621600 T_2^2
T_6 T_1^3+43243200 T_4 T_6 T_1^3 
+49420800 T_3 T_7
   T_1^3+64864800 T_2 T_8 T_1^3-103783680 T_{10} T_1^3
\nonumber\\&&
-9609600 T_2 T_3^3 T_1^2+32432400 T_3 T_4^2 T_1^2+2702700 T_2^4 T_3
   T_1^2-32432400 T_2^2 T_3 T_4 T_1^2
\nonumber\\&&
-12972960 T_2^3 T_5 T_1^2+34594560 T_3^2 T_5 T_1^2+77837760 T_2 T_4
T_5 T_1^2+86486400 
   T_2 T_3 T_6 T_1^2
\nonumber\\&&
-103783680 T_5 T_6 T_1^2+55598400 T_2^2 T_7 T_1^2-111196800 T_4 T_7
T_1^2-129729600 T_3 T_8 
   T_1^2
\nonumber\\&&
-172972800 T_2 T_9 T_1^2+283046400 T_{11} T_1^2+135135 T_2^6
T_1+3203200 T_3^4 T_1-16216200 T_4^3 T_1 
\nonumber\\&&
-7207200 T_2^3
   T_3^2 T_1+24324300 T_2^2 T_4^2 T_1-62270208 T_2 T_5^2 T_1+86486400 T_6^2 T_1
\nonumber\\&&
-4054050 T_2^4 T_4 T_1+43243200 T_2 T_3^2 T_4
   T_1+51891840 T_2^2 T_3 T_5 T_1-103783680 T_3 T_4 T_5 T_1
\nonumber\\&&
+21621600 T_2^3 T_6 T_1-57657600 T_3^2 T_6 T_1-129729600 T_2 T_4
   T_6 T_1-148262400 T_2 T_3 T_7 T_1
\nonumber\\&&
+177914880 T_5 T_7 T_1-97297200 T_2^2 T_8 T_1+194594400 T_4 T_8 T_1+230630400 T_3 T_9 T_1
\nonumber\\&&
+311351040 T_2 T_{10} T_1-518918400 T_{12} T_1+4804800 T_2^2 T_3^3-32432400 T_2 T_3 T_4^2
\nonumber\\&&
+41513472 T_3 T_5^2-540540
   T_2^5 T_3-9609600 T_3^3 T_4+10810800 T_2^3 T_3 T_4
\nonumber\\&&
+3243240 T_2^4 T_5-34594560 T_2 T_3^2 T_5+38918880 T_4^2 T_5-38918880
   T_2^2 T_4 T_5
\nonumber\\&&
-43243200 T_2^2 T_3 T_6+86486400 T_3 T_4 T_6+103783680 T_2 T_5 T_6-18532800 T_2^3 T_7
\nonumber\\&&
+49420800 T_3^2
   T_7+111196800 T_2 T_4 T_7-148262400 T_6 T_7+129729600 T_2 T_3 T_8
\nonumber\\&&
-155675520 T_5 T_8+86486400 T_2^2 T_9-172972800 T_4
   T_9-207567360 T_3 T_{10}
\nonumber\\&&
-283046400 T_2 T_{11}+479001600 T_{13}\,.
\end{eqnarray}
}

\section{Numerical Precision in $n$-Meson Correlation Functions}
\label{sec:numer-prec-n}

\noindent
In order to extract the correlation functions for systems with $n\gsim
8$ mesons it is necessary to perform the contractions using high
precision numerical techniques. It is necessary to calculate the
propagators to an analogous level of precision.

\subsection{Contraction Precision}
\label{sec:contr-prec}

\noindent
As discussed in the main text, the double precision numerical
representations are insufficient to accurately compute the propagator
contractions required for the large $n$ correlators. Here we exhibit
the failure of these calculations and discuss their origin.
\begin{figure}[!t]
  \centering
  \includegraphics[width=0.90\columnwidth]{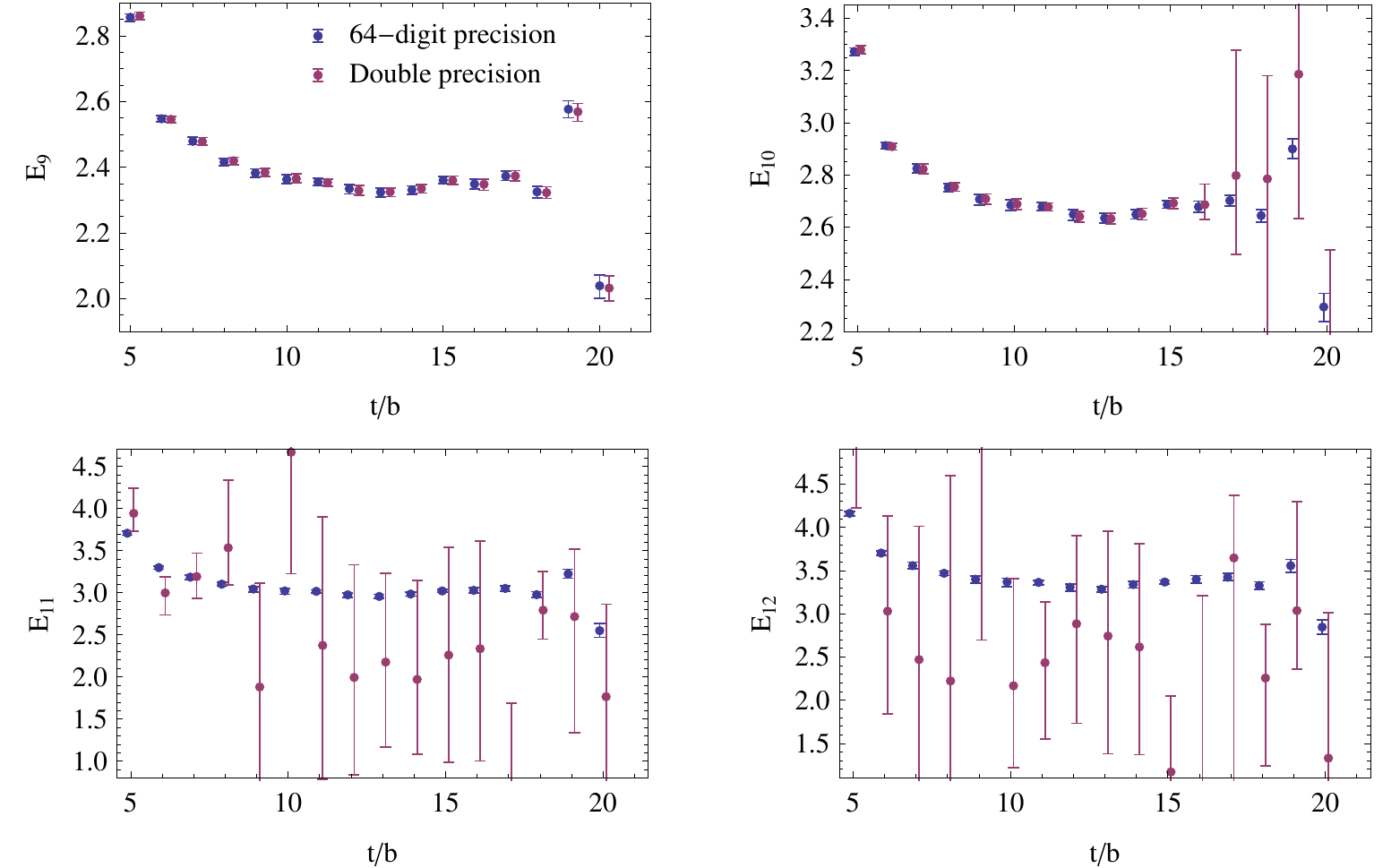}
  \caption{Comparison of double precision contractions and 64 decimal digit precision
    contractions for the effective energy plots for $n=9,\ 10,\ 11$
    and 12 pions.  }
\label{fig:comparison010}
\end{figure}
\begin{figure}[!t]
  \centering
  \includegraphics[width=0.90\columnwidth]{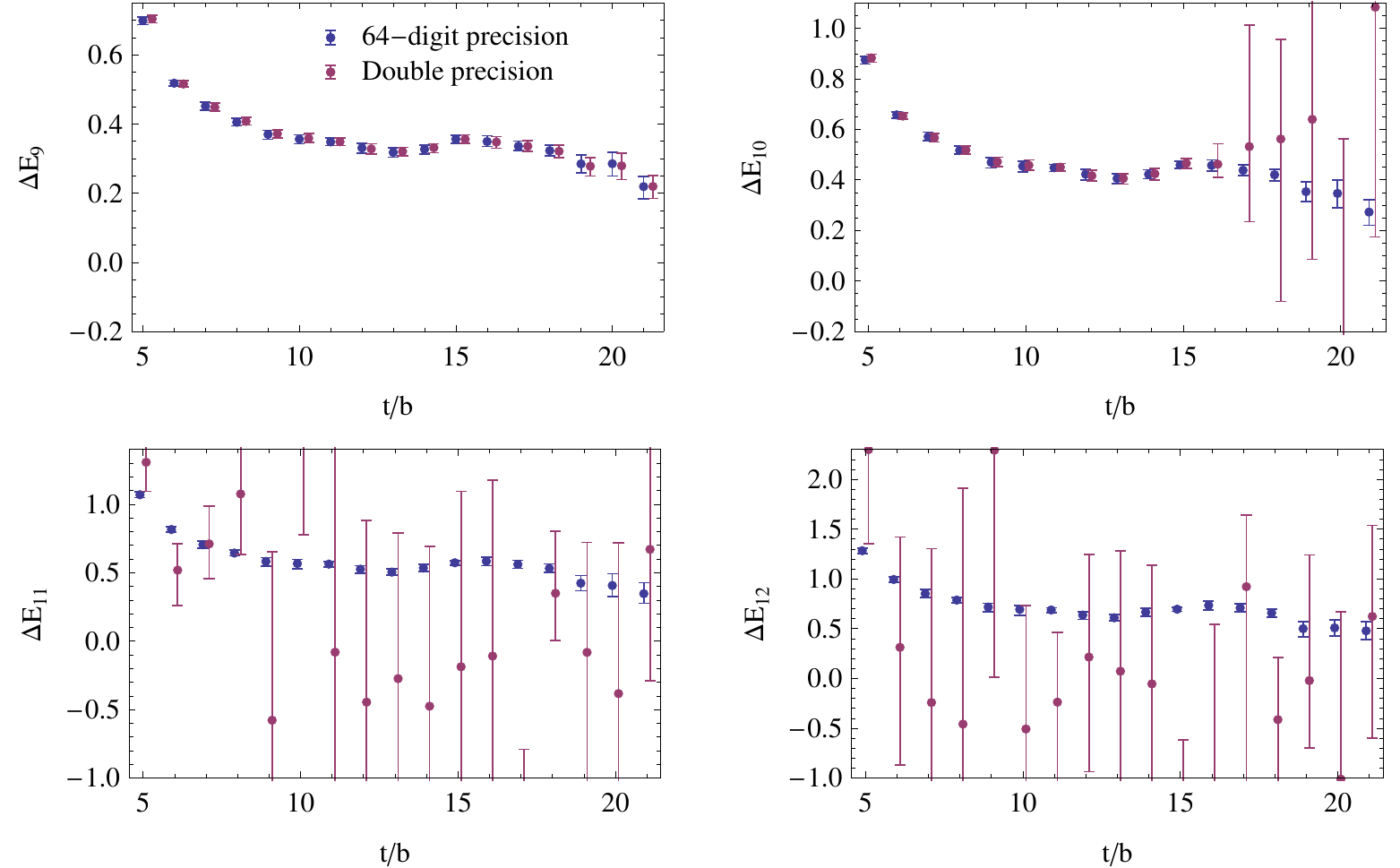}
  \caption{Comparison of double precision contractions and 64 decimal digit precision
    contractions for the effective energy difference plots for $n=9,\ 
    10,\ 11$ and 12 pions.  }
\label{fig:deltacomparison010}
\end{figure}
In figs.~\ref{fig:comparison010} and \ref{fig:deltacomparison010}, we
compare the $n$ particle effective energies and effective energy
differences performed using double precision (squares) and using 64
decimal digit internal precision (stars, slightly offset to the right
for clarity) for $n=9$, 10, 11 and 12 for the $m_\pi=352$~MeV
ensemble.  The breakdown of double-precision calculations with
increasing $n$ and $t$, where non-zero values of the correlation
functions are lost to noise, is clear.  This breakdown has its origin
in the large powers to which propagators are raised in the
contractions required for systems with a large number of mesons.
Viewing the zero three-momentum propagator as a time-dependent
$12\times12$ matrix, the range of numbers that can be represented in
double precision is too limited and small elements of high powers of
these propagators are rounded away in tracing or other matrix
manipulations.  Such terms are crucial for maintaining gauge
invariance, and their removal leads to a serious degradation of the
correlation function.  For the numbers of mesons considered in this
work, $n\le 13$, it is sufficient to use 64 decimal digit precision,
but for $n\gsim 13$, such precision will rapidly become insufficient.

\subsection{Propagator Precision}
\label{sec:propagator-precision}

\noindent
Given a particular gauge field, the accuracy with which the quark
propagator is computed (e.g. the tolerance in conjugate gradient (CG)
algorithm) will influence the correlation functions computed from that
propagator. In extreme cases, where the solution is quite inaccurate
in comparison to the statistical precision of the importance sampling
procedure, this residual error can persist through the ensemble
averaging and lead to errors in the correlation functions.  In the
case of the multi-pion correlations studied here, the situation is
particularly acute as the high powers to which the propagators are
taken can enhance small effects.  At this point in time we have not
performed a detailed study of this issue.  It would require generating
full sets of propagators at different inversion precisions on the same
configurations.  We have simply studied the effect on a few
representative configurations, and it is clear that for large $n$,
significant differences in the correlation functions can arise from
loose tolerance of the CG-solver.  In future studies of even larger
$n$ (and very high precision studies of any observable), this issue
must be investigated in more detail.

%
%

\end{document}